\documentclass[a4paper,11pt]{article}
\usepackage{jheppub} 
\usepackage[T1]{fontenc} 
\usepackage{amsmath,amsthm}
\usepackage{amssymb, amsfonts}
\usepackage{bbm}

\usepackage{graphicx}
\usepackage{dsfont}
\usepackage{relsize}
\usepackage{stmaryrd}
\usepackage{lmodern}
\usepackage{slantsc}
\usepackage{scalefnt}
\usepackage{wasysym}
\usepackage{xspace}
\usepackage{boxedminipage}
\usepackage{ifpdf}
\usepackage{multirow, hhline}
\usepackage{slashed}
\usepackage{xifthen}
\usepackage{tensor}
\usepackage{array, nicematrix}
\usepackage{tabu}
\usepackage{pbox}
\usepackage{multirow}
\usepackage{tikz}
\usepackage{overpic}
\usepackage{color}
\usepackage{diagbox}
\usepackage{makecell}
\usetikzlibrary{arrows, matrix, calc, scopes, decorations.markings}
\usepackage[mathcal]{euscript}
\usepackage{booktabs}
\usepackage{autobreak, subfig, diagbox, mathrsfs}
\usepackage{mathbbol}
\usepackage{tcolorbox}
\usepackage{hyperref}
\allowdisplaybreaks[4]

\newcommand{\NN}{\mathbb{N}}

\newcommand{\A}{\mathcal{A}}
\newcommand{\Hil}{\mathcal{H}}

\newcommand{\T}{\mathcal{T}}
\newcommand{\Bimod}{{\mathsf{Bimod}}}
\newcommand{\Fibo}{{\mathsf{Fibo}}}
\newcommand{\Child}{{\mathcal{S}}}
\newcommand{\C}{\mathbb{C}}

\newcommand{\D}{\mathscr{D}}

\newcommand{\Fus}{\mathcal{F}}
\newcommand{\Sub}{\mathcal{S}}
\newcommand{\Cent}{\mathcal{Z}}
\newcommand{\F}{\mathscr{F}}
\newcommand{\RR}{\mathbb{R}}
\newcommand{\CC}{\mathbb{C}}
\newcommand{\idm}{{\mathds{1}}}
\newcommand{\trivial}{{\mathbf{1}}}

\newcommand{\df}{{\mathrm{d}}}

\newcommand{\dk}[2][1]{{\ifthenelse{\equal{#1}{1}}{\frac{\df{#2}}{2\pi}}{\frac{\df^{#1}{#2}}{(2\pi)^{#1}}}}}

\newcommand{\rep}{{\sf Rep}}

\newcommand{\name}[1]{{\textbf{\textit{#1}}}}

\newcommand{\ket}[1]{{\left\vert{#1}\right\rangle}}

\newcommand{\eq}[1]{\begin{align*}#1\end{align*}}
\newcommand{\eqn}[2][0]{\ifthenelse{\equal{#1}{0}}{\begin{equation}\begin{aligned}#2\end{aligned}\end{equation}}{\begin{equation}\begin{aligned}#2\end{aligned}\label{#1}\end{equation}}}

\usetikzlibrary{shapes.geometric, snakes}
\tikzset{>=latex}
\usetikzlibrary{arrows, matrix, calc, scopes, decorations.markings}
\usetikzlibrary{decorations.pathmorphing, positioning}
\tikzset{snake it/.style={decorate, decoration={snake,amplitude=0.2mm,segment length=1mm}}}
\tikzset{->-/.style={decoration={
			 markings,
			 mark=at position .5*\pgfdecoratedpathlength+2pt with {\arrow{>}}},postaction={decorate}}}

\tikzset{-<-/.style={decoration={
			 markings,
			 mark=at position .5*\pgfdecoratedpathlength+2pt with {\arrow{<}}},postaction={decorate}}}

\newtabulinestyle{dash=off 2pt}

{\null\,\vcenter\bgroup\scriptsize
\arraycolsep=.13885em
\hbox\bgroup$\array{@{}#1@{}}}
{\endarray$\egroup\egroup\,\null}
\newcommand{\Sketch}{
\begin{tikzpicture}[baseline={([yshift=-.8ex]current bounding box.center)},scale=1]
\fill [gray!15!, rounded corners] (-1.3, 1.) -- (-1.3, -1.4) -- (6.5, -1.4) -- (6.5, 3.4) -- (-1.3, 3.4) -- (-1.3, 1.);
\fill [gray!40!, rounded corners] (-1, 1.) -- (-1., .4) -- (5., .4) -- (5., 1.6) -- (-1., 1.6) -- (-1, 1.);
\fill [gray!50!, rounded corners] (-1, 2.) -- (-1., 1.7) -- (5., 1.7) -- (5., 3.) -- (-1., 3.) -- (-1, 2.);
\fill [gray!50!, rounded corners] (-1, 0) -- (-1., -1) -- (5., -1) -- (5., .3) -- (-1., .3) -- (-1, 0);
\fill [gray!15!, rounded corners] (7.3, 1.) -- (7.3, -1.4) -- (13.7, -1.4) -- (13.7, 3.4) -- (7.3, 3.4) -- (7.3, 1.);
\fill [gray!50!, rounded corners] (7.6, 2.) -- (7.6, 1.7) -- (13.4, 1.7) -- (13.4, 3.) -- (7.6, 3.) -- (7.6, 2.);
\fill [gray!50!, rounded corners] (7.6, 0) -- (7.6, -1) -- (13.4, -1) -- (13.4, .3) -- (7.6, .3) -- (7.6, 0);
\node [centered, rotate=0.] at (0, 2) {\scriptsize Parent Model};
\node [centered, rotate=0.] at (0, 0) {\scriptsize Equivalent};
\node [centered, rotate=0.] at (0, -.3) {\scriptsize Parent Model};
\node [centered, rotate=0.] at (4, 0) {\scriptsize Equivalent};
\node [centered, rotate=0.] at (4, -.3) {\scriptsize Child Model};
\node [centered, rotate=0.] at (4, 2) {\scriptsize Child Model};

\draw [red, thick, ->] (0, 1.6) -- (0, .4);
\node [red, centered, rotate=0.] at (-.3, 1.) {\scriptsize $\D^\dagger$};
\draw [red, thick, ->] (4, .4) -- (4, 1.6);
\node [red, centered, rotate=0.] at (4.3, 1.) {\scriptsize $\D$};
\draw [->] (1.1, -.15) -- (3, -.15);
\node [centered, rotate=0.] at (2., -.35) {\scriptsize Fluxon};
\node [centered, rotate=0.] at (2., -.65) {\scriptsize Condensation $P$};
\draw [->] (1.1, 2.) -- (3, 2.);
\node [centered, rotate=0.] at (2., 2.8) {\scriptsize Non-Fluxon};
\node [centered, rotate=0.] at (2., 2.5) {\scriptsize Condensation};
\node [centered, rotate=0.] at (2., 2.2) {\scriptsize $\D P\D^\dagger$};
\fill [red] (2, 0.4) -- (1.95, 1.43) -- (1.85, 1.35) -- (2, 1.6) -- (2.15, 1.35) -- (2.05, 1.43) -- (2, 0.4);

\node [red, centered, rotate=0.] at (5.5, 1.1) {\scriptsize\it Our};
\node [red, centered, rotate=0.] at (5.5, .7) {\scriptsize\it Recipe};

\node [centered, rotate=0.] at (8.6, 2.3) {\scriptsize Parent Input};
\node [centered, rotate=0.] at (8.6, 1.95) {\scriptsize UFC $\Fus$};
\node [centered, rotate=0.] at (8.6, 0.) {\scriptsize Equivalent};
\node [centered, rotate=0.] at (8.6, -0.35) {\scriptsize Input UFC $\Fus'$};
\node [centered, rotate=0.] at (12.4, 0.) {\scriptsize Equivalent};
\node [centered, rotate=0.] at (12.4, -.35) {\scriptsize Child Input};
\node [centered, rotate=0.] at (12.4, -.7) {\scriptsize UFC $\Sub'\subset \Fus'$};
\node [centered, rotate=0.] at (12.4, 2.3) {\scriptsize Child Input};
\node [centered, rotate=0.] at (12.4, 1.95) {\scriptsize UFC $\Sub\subset\Fus$};

\draw [->] (9.6, -.15) -- (11.5, -.15);
\node [centered, rotate=0.] at (10.5, -.35) {\scriptsize Select a Full};
\node [centered, rotate=0.] at (10.5, -.65) {\scriptsize Subcategory};
\draw [->] (9.55, 2.2) -- (11.55, 2.2);
\node [centered, rotate=0.] at (10.5, 2.7) {\scriptsize Select a Certain};
\node [centered, rotate=0.] at (10.5, 2.4) {\scriptsize Subcategory};
\draw [red, thick] (10.4, 0.3) -- (10.4, 1.6);
\draw [red, thick] (10.5, 0.3) -- (10.5, 1.6);
\draw [red, thick] (10.28, 1.44) -- (10.45, 1.68) -- (10.62, 1.44);
\node [red, right, rotate=0.] at (10.4, 1.1) {\scriptsize\it Pseudonatural};
\node [red, right, rotate=0.] at (10.4, .7) {\scriptsize\it Transformation};

\fill [blue] (6., .85) -- (10., .85) -- (10., .75) -- (10.3, .9) -- (10., 1.05) -- (10., .95) -- (6., .95) -- (6., .85);
\node [centered, rotate=0., blue] at (8.2, 1.25) {\it Represents};
\draw [->, dashed, blue] (3., 2.6) -- (4.7, 3.8);
\draw [->, dashed, blue] (9.3, 2.8) -- (8.5, 3.8);
\node [centered, rotate=0., blue] at (6.6, 4.) {\it Previously Unknown!};
\node [centered, rotate=0.] at (2., -1.7) {\scriptsize (a) Our Model-Based Realization};
\node [centered, rotate=0.] at (10.4, -1.7) {\scriptsize (b) A Finer Categorical Framework};
\end{tikzpicture}
}

\newcommand{\Pseudo}{
\begin{tikzpicture}[baseline={([yshift=-.8ex]current bounding box.center)},scale=1]
\node [centered, rotate=0.] at (0, 2.) {\scriptsize $\Fus$};
\node [centered, rotate=0.] at (0, 0.) {\scriptsize $\Bimod_\Fus(\A)$};
\node [centered, rotate=0.] at (5, 0.) {\scriptsize $\Sub'$};
\node [centered, rotate=0.] at (5, 2.) {\scriptsize $\Child$};
\draw [<-] (.9, 0) -- (4.8, 0);
\node [centered, rotate=0.] at (2.85, .25) {\scriptsize Full Subcategory Injection};
\draw [<-] (.3, 2) -- (4.8, 2);
\node [centered, rotate=0.] at (2.5, 2.25) {\scriptsize Subcategory Injection};
\draw [red, ultra thick, ->] (0, .2) -- (0, 1.8);
\draw [red, very thick] (2.49, .5) -- (2.49, 1.7);
\draw [red, very thick] (2.41, .5) -- (2.41, 1.7);
\fill [red] (2.45, 1.85) -- (2.3, 1.5) -- (2.45, 1.65) -- (2.6, 1.5);
\draw [red, ultra thick, ->] (5, .2) -- (5, 1.8);
\node [red, centered, rotate=0.] at (2.5, 1.) {\scriptsize\it Pesudonatural\ \ \ \ Transformation};
\end{tikzpicture}
}

\newcommand{\Lattice}{
\begin{tikzpicture}[baseline={([yshift=-.8ex]current bounding box.center)},scale=1]
\draw [decorate,decoration={snake,segment length=.12cm, amplitude=.04cm}, ->] (1.7,6.8) -- (2.3,6.8);
\draw [->-] (1.7,6.4) -- (1.7,6.8);
\draw [->-] (1.7,6.8) -- (1.7,7.2);
\draw [->-] (1.7,7.2) -- (2.4,7.6);
\draw [->-] (2.4,7.6) -- (3.1,7.2);
\draw [->-] (2.4,6.) -- (1.7,6.4);
\draw [->-] (3.1,6.4) -- (2.4,6.);
\draw [->-] (2.4,5.2) -- (2.4,5.6);
\draw [->-] (2.4,5.6) -- (2.4,6.);
\draw [decorate,decoration={snake,segment length=.12cm, amplitude=.04cm}, ->] (2.4,5.6) -- (3.,5.6);
\draw [decorate,decoration={snake,segment length=.12cm, amplitude=.04cm}, ->] (3.1,6.8) -- (3.7,6.8);
\draw [->-] (3.1,6.4) -- (3.1,6.8);
\draw [->-] (3.1,6.8) -- (3.1,7.2);
\draw [->-] (3.1,7.2) -- (3.8,7.6);
\draw [->-] (3.8,7.6) -- (4.5,7.2);
\draw [->-] (3.8,6.) -- (3.1,6.4);
\draw [->-] (4.5,6.4) -- (3.8,6.);
\draw [->-] (3.8,5.2) -- (3.8,5.6);
\draw [->-] (3.8,5.6) -- (3.8,6.);
\draw [decorate,decoration={snake,segment length=.12cm, amplitude=.04cm}, ->] (3.8,5.6) -- (4.4,5.6);
\draw [decorate,decoration={snake,segment length=.12cm, amplitude=.04cm}, ->] (4.5,6.8) -- (5.1,6.8);
\draw [->-] (4.5,6.4) -- (4.5,6.8);
\draw [->-] (4.5,6.8) -- (4.5,7.2);
\draw [->-] (4.5,7.2) -- (5.2,7.6);
\draw [->-] (5.2,7.6) -- (5.9,7.2);
\draw [->-] (5.2,6.) -- (4.5,6.4);
\draw [->-] (5.9,6.4) -- (5.2,6.);
\draw [->-] (5.2,5.2) -- (5.2,5.6);
\draw [->-] (5.2,5.6) -- (5.2,6.);
\draw [decorate,decoration={snake,segment length=.12cm, amplitude=.04cm}, ->] (5.2,5.6) -- (5.8,5.6);
\draw [decorate,decoration={snake,segment length=.12cm, amplitude=.04cm}, ->] (5.9,6.8) -- (6.5,6.8);
\draw [->-] (5.9,6.4) -- (5.9,6.8);
\draw [->-] (5.9,6.8) -- (5.9,7.2);
\draw [->-] (5.9,7.2) -- (6.6,7.6);
\draw [->-] (6.6,7.6) -- (7.3,7.2);
\draw [->-] (6.6,6.) -- (5.9,6.4);
\draw [->-] (7.3,6.4) -- (6.6,6.);
\draw [->-] (6.6,5.2) -- (6.6,5.6);
\draw [->-] (6.6,5.6) -- (6.6,6.);
\draw [decorate,decoration={snake,segment length=.12cm, amplitude=.04cm}, ->] (6.6,5.6) -- (7.2,5.6);
\draw [decorate,decoration={snake,segment length=.12cm, amplitude=.04cm}, ->] (1.7,4.4) -- (2.3,4.4);
\draw [->-] (1.7,4.) -- (1.7,4.4);
\draw [->-] (1.7,4.4) -- (1.7,4.8);
\draw [->-] (1.7,4.8) -- (2.4,5.2);
\draw [->-] (2.4,5.2) -- (3.1,4.8);
\draw [->-] (2.4,3.6) -- (1.7,4.);
\draw [->-] (3.1,4.) -- (2.4,3.6);
\draw [->-] (2.4,3.2) -- (2.4,3.6);
\draw [decorate,decoration={snake,segment length=.12cm, amplitude=.04cm}, ->] (3.1,4.4) -- (3.7,4.4);
\draw [->-] (3.1,4.) -- (3.1,4.4);
\draw [->-] (3.1,4.4) -- (3.1,4.8);
\draw [->-] (3.1,4.8) -- (3.8,5.2);
\draw [->-] (3.8,5.2) -- (4.5,4.8);
\draw [->-] (3.8,3.6) -- (3.1,4.);
\draw [->-] (4.5,4.) -- (3.8,3.6);
\draw [->-] (3.8,3.2) -- (3.8,3.6);
\draw [decorate,decoration={snake,segment length=.12cm, amplitude=.04cm}, ->] (4.5,4.4) -- (5.1,4.4);
\draw [->-] (4.5,4.) -- (4.5,4.4);
\draw [->-] (4.5,4.4) -- (4.5,4.8);
\draw [->-] (4.5,4.8) -- (5.2,5.2);
\draw [->-] (5.2,5.2) -- (5.9,4.8);
\draw [->-] (5.2,3.6) -- (4.5,4.);
\draw [->-] (5.9,4.) -- (5.2,3.6);
\draw [->-] (5.2,3.2) -- (5.2,3.6);
\draw [decorate,decoration={snake,segment length=.12cm, amplitude=.04cm}, ->] (5.9,4.4) -- (6.5,4.4);
\draw [->-] (5.9,4.) -- (5.9,4.4);
\draw [->-] (5.9,4.4) -- (5.9,4.8);
\draw [->-] (5.9,4.8) -- (6.6,5.2);
\draw [->-] (6.6,5.2) -- (7.3,4.8);
\draw [->-] (6.6,3.6) -- (5.9,4.);
\draw [->-] (7.3,4.) -- (6.6,3.6);
\draw [->-] (6.6,3.2) -- (6.6,3.6);
\draw [decorate,decoration={snake,segment length=.12cm, amplitude=.04cm}, ->] (7.3,6.8) -- (7.9,6.8);
\draw [->-] (7.3,6.4) -- (7.3,6.8);
\draw [->-] (7.3,6.8) -- (7.3,7.2);
\draw [->-] (7.3,7.2) -- (7.6,7.4);
\draw [->-] (8.,6.) -- (7.3,6.4);
\draw [->-] (8.3,6.2) -- (8.,6.);
\draw [->-] (8.,5.2) -- (8.,5.6);
\draw [->-] (8.,5.6) -- (8.,6.);
\draw [decorate,decoration={snake,segment length=.12cm, amplitude=.04cm}, ->] (8.,5.6) -- (8.6,5.6);
\draw [decorate,decoration={snake,segment length=.12cm, amplitude=.04cm}, ->] (7.3,4.4) -- (7.9,4.4);
\draw [->-] (7.3,4.) -- (7.3,4.4);
\draw [->-] (7.3,4.4) -- (7.3,4.8);
\draw [->-] (7.3,4.8) -- (8.,5.2);
\draw [->-] (8.,5.2) -- (8.3,5.);
\draw [->-] (8.,3.6) -- (7.3,4.);
\draw [->-] (8.,3.2) -- (8.,3.6);
\draw [->-] (6.6,7.6) -- (6.6,8.);
\draw [->-] (5.2,7.6) -- (5.2,8.);
\draw [->-] (3.8,7.6) -- (3.8,8.);
\draw [->-] (2.4,7.6) -- (2.4,8.);
\draw [->-] (8.3,3.8) -- (8.,3.6);
\draw [->-] (1.7,4.) -- (1.3,3.8);
\draw [->-] (1.7,4.8) -- (1.3,5.);
\draw [->-] (1.7,6.4) -- (1.3,6.2);
\draw [->-] (1.7,7.2) -- (1.3,7.4);
\end{tikzpicture}
}

\newcommand{\PlaquetteSrc}{
\begin{tikzpicture}[baseline={([yshift=-.8ex]current bounding box.center)},scale=.5]
\draw [->-] (3.2,5.6) -- (3.2,6.8);
\draw [->-] (3.2,7.2) -- (4.6,8.);
\draw [->-] (4.6,8.) -- (6.,7.2);
\draw [->-] (6.,5.6) -- (4.6,4.8);
\draw [->-] (4.6,4.8) -- (3.2,5.6);
\draw [->-] (6.,7.2) -- (6.,5.6);
\draw [decorate,decoration={snake,segment length=.12cm, amplitude=.04cm}, ->] (3.2,6.4) -- (4.4,6.4);
\draw [->-] (2.6,7.6) -- (3.2,7.2);
\draw [->-] (4.6,8.8) -- (4.6,8.);
\draw [->-] (6.6,7.6) -- (6.,7.2);
\draw [->-] (6.6,5.2) -- (6.,5.6);
\draw [->-] (4.6,4.) -- (4.6,4.8);
\draw [->-] (2.6,5.2) -- (3.2,5.6);
\draw [->-] (3.2,6.6) -- (3.2,7.2);
\node [centered, rotate=0.] at (2.4,7.8) {\scriptsize $e_1$};
\node [centered, rotate=0.] at (4.95,8.8) {\scriptsize $e_2$};
\node [centered, rotate=0.] at (6.9,7.8) {\scriptsize $e_3$};
\node [centered, rotate=0.] at (6.9,5.) {\scriptsize $e_4$};
\node [centered, rotate=0.] at (5.,4.) {\scriptsize $e_5$};
\node [centered, rotate=0.] at (2.3,5.) {\scriptsize $e_6$};
\node [centered, rotate=0.] at (2.85,6.) {\scriptsize $i_7$};
\node [centered, rotate=0.] at (2.85,6.9) {\scriptsize $i_1$};
\node [centered, rotate=0.] at (3.6,7.9) {\scriptsize $i_2$};
\node [centered, rotate=0.] at (5.6,7.9) {\scriptsize $i_3$};
\node [centered, rotate=0.] at (6.4,6.4) {\scriptsize $i_4$};
\node [centered, rotate=0.] at (5.6,4.9) {\scriptsize $i_5$};
\node [centered, rotate=0.] at (3.7,4.9) {\scriptsize $i_6$};
\node [centered, rotate=0.] at (4.7,6.4) {\scriptsize $p$};
\end{tikzpicture}
}

\newcommand{\PlaquetteTar}{
\begin{tikzpicture}[baseline={([yshift=-.8ex]current bounding box.center)},scale=.5]
\draw [->-] (3.2,5.6) -- (3.2,6.8);
\draw [->-] (3.2,7.2) -- (4.6,8.);
\draw [->-] (4.6,8.) -- (6.,7.2);
\draw [->-] (6.,5.6) -- (4.6,4.8);
\draw [->-] (4.6,4.8) -- (3.2,5.6);
\draw [->-] (6.,7.2) -- (6.,5.6);
\draw [decorate,decoration={snake,segment length=.12cm, amplitude=.04cm}, ->] (3.2,6.4) -- (4.4,6.4);
\draw [->-] (2.6,7.6) -- (3.2,7.2);
\draw [->-] (4.6,8.8) -- (4.6,8.);
\draw [->-] (6.6,7.6) -- (6.,7.2);
\draw [->-] (6.6,5.2) -- (6.,5.6);
\draw [->-] (4.6,4.) -- (4.6,4.8);
\draw [->-] (2.6,5.2) -- (3.2,5.6);
\draw [->-] (3.2,6.6) -- (3.2,7.2);
\node [centered, rotate=0.] at (2.4,7.8) {\scriptsize $e_1$};
\node [centered, rotate=0.] at (4.95,8.8) {\scriptsize $e_2$};
\node [centered, rotate=0.] at (6.9,7.8) {\scriptsize $e_3$};
\node [centered, rotate=0.] at (6.9,5.) {\scriptsize $e_4$};
\node [centered, rotate=0.] at (5.,4.) {\scriptsize $e_5$};
\node [centered, rotate=0.] at (2.3,5.) {\scriptsize $e_6$};
\node [centered, rotate=0.] at (2.85,6.) {\scriptsize $j_7$};
\node [centered, rotate=0.] at (2.85,6.9) {\scriptsize $j_1$};
\node [centered, rotate=0.] at (3.6,7.9) {\scriptsize $j_2$};
\node [centered, rotate=0.] at (5.6,7.9) {\scriptsize $j_3$};
\node [centered, rotate=0.] at (6.4,6.4) {\scriptsize $j_4$};
\node [centered, rotate=0.] at (5.6,4.9) {\scriptsize $j_5$};
\node [centered, rotate=0.] at (3.7,4.9) {\scriptsize $j_6$};
\node [centered, rotate=0.] at (4.7,6.4) {\scriptsize $1$};
\end{tikzpicture}
}

\newcommand{\ExcitedSrc}{
\begin{tikzpicture}[baseline={([yshift=-.8ex]current bounding box.center)},scale=.8]
\draw [->-] (2.,2.9) -- (1.3,3.3);
\draw [->-] (1.3,3.3) -- (0.6,2.9);
\draw [->-] (0.6,2.9) -- (0.6,2.);
\draw [->-] (0.6,2.) -- (1.3,1.6);
\draw [->-] (2.,2.9) -- (2.7,3.3);
\draw [->-] (2.7,3.3) -- (3.4,2.9);
\draw [->-] (3.4,2.9) -- (3.4,2.);
\draw [->-] (3.4,2.) -- (2.7,1.6);
\draw [->-] (1.3,1.6) -- (2.,2.);
\draw [->-] (2.7,1.6) -- (2.,2.);
\draw [->-] (1.3,3.3) -- (1.3,3.7);
\draw [->-] (2.7,3.3) -- (2.7,3.7);
\draw [->-] (3.4,2.9) -- (3.7,3.1);
\draw [->-] (3.4,2.) -- (3.7,1.8);
\draw [->-] (2.7,1.6) -- (2.7,1.2);
\draw [->-] (1.3,1.6) -- (1.3,1.2);
\draw [->-] (0.6,2.) -- (0.3,1.8);
\draw [->-] (0.6,2.9) -- (0.3,3.1);
\draw [->-] (2, 2.1) -- (2, 2.9);
\draw [->-] (2, 2) -- (2, 2.3);
\draw [->-] (2, 2.7) -- (2, 2.9);
\draw [dashed] (2, 2.25) -- (1.5, 2.25);
\draw [dashed] (2, 2.65) -- (2.5, 2.65);
\node [centered, rotate=0.] at (2.35,2.42) {\tiny $j_E$};
\node [centered, rotate=0.] at (1.3,2.2) {\tiny $1$};
\node [centered, rotate=0.] at (2.7,2.7) {\tiny $1$};
\end{tikzpicture}
}

\newcommand{\ExcitedTar}{
\begin{tikzpicture}[baseline={([yshift=-.8ex]current bounding box.center)},scale=.8]
\draw [->-] (2.,2.9) -- (1.3,3.3);
\draw [->-] (1.3,3.3) -- (0.6,2.9);
\draw [->-] (0.6,2.9) -- (0.6,2.);
\draw [->-] (0.6,2.) -- (1.3,1.6);
\draw [->-] (2.,2.9) -- (2.7,3.3);
\draw [->-] (2.7,3.3) -- (3.4,2.9);
\draw [->-] (3.4,2.9) -- (3.4,2.);
\draw [->-] (3.4,2.) -- (2.7,1.6);
\draw [->-] (1.3,1.6) -- (2.,2.);
\draw [->-] (2.7,1.6) -- (2.,2.);
\draw [->-] (1.3,3.3) -- (1.3,3.7);
\draw [->-] (2.7,3.3) -- (2.7,3.7);
\draw [->-] (3.4,2.9) -- (3.7,3.1);
\draw [->-] (3.4,2.) -- (3.7,1.8);
\draw [->-] (2.7,1.6) -- (2.7,1.2);
\draw [->-] (1.3,1.6) -- (1.3,1.2);
\draw [->-] (0.6,2.) -- (0.3,1.8);
\draw [->-] (0.6,2.9) -- (0.3,3.1);

\draw [->-] (2, 2) -- (2, 2.3);
\draw [->-] (2, 2.3) -- (2, 2.6);
\draw [->-] (2, 2.6) -- (2, 2.9);
\draw [decorate,decoration={snake,segment length=.12cm, amplitude=.04cm}, ->] (2.,2.3) -- (1.5,2.3);
\draw [decorate,decoration={snake,segment length=.12cm, amplitude=.04cm}, ->] (2.,2.6) -- (2.5,2.6);
\node [centered, rotate=0.] at (2.25,2.15) {\tiny $j_E$};
\node [centered, rotate=0.] at (2.13,2.42) {\tiny $k$};
\node [centered, rotate=0.] at (1.75,2.77) {\tiny $j_E$};
\node [centered, rotate=0.] at (1.35,2.35) {\scriptsize $p^\ast$};
\node [centered, rotate=0.] at (2.6,2.6) {\scriptsize $q$};
\end{tikzpicture}
}

\newcommand{\ExcitedFlux}{
\begin{tikzpicture}[baseline={([yshift=-.8ex]current bounding box.center)},scale=.8]
\draw [->-] (2.,2.9) -- (1.3,3.3);
\draw [->-] (1.3,3.3) -- (0.6,2.9);
\draw [->-] (0.6,2.9) -- (0.6,2.);
\draw [->-] (0.6,2.) -- (1.3,1.6);
\draw [->-] (2.,2.9) -- (2.7,3.3);
\draw [->-] (2.7,3.3) -- (3.4,2.9);
\draw [->-] (3.4,2.9) -- (3.4,2.);
\draw [->-] (3.4,2.) -- (2.7,1.6);
\draw [->-] (1.3,1.6) -- (2.,2.);
\draw [->-] (2.7,1.6) -- (2.,2.);
\draw [->-] (1.3,3.3) -- (1.3,3.7);
\draw [->-] (2.7,3.3) -- (2.7,3.7);
\draw [->-] (3.4,2.9) -- (3.7,3.1);
\draw [->-] (3.4,2.) -- (3.7,1.8);
\draw [->-] (2.7,1.6) -- (2.7,1.2);
\draw [->-] (1.3,1.6) -- (1.3,1.2);
\draw [->-] (0.6,2.) -- (0.3,1.8);
\draw [->-] (0.6,2.9) -- (0.3,3.1);
\draw [->-] (2, 2.1) -- (2, 2.9);
\draw [->-] (2, 2) -- (2, 2.3);
\draw [->-] (2, 2.7) -- (2, 2.9);
\draw [decorate,decoration={snake,segment length=.12cm, amplitude=.04cm}, ->]  (2, 2.25) -- (1.5, 2.25);
\draw [decorate,decoration={snake,segment length=.12cm, amplitude=.04cm}, ->] (2, 2.6) -- (2.5, 2.6);
\node [centered, rotate=0.] at (2.16,2.08) {\tiny $i$};
\node [centered, rotate=0.] at (2.28,2.4) {\tiny $j_E$};
\node [centered, rotate=0.] at (1.77,2.8) {\tiny $k$};
\node [centered, rotate=0.] at (1.3,2.2) {\tiny $p$};
\node [centered, rotate=0.] at (2.65,2.6) {\tiny $q$};
\end{tikzpicture}
}

\newcommand{\Edge}[1]{
\begin{tikzpicture}[baseline={([yshift=-.8ex]current bounding box.center)},scale=1.]
\draw [] (0, 0) -- (0, 1.5);
\node [right, rotate=0.] at (0.,.75) {\scriptsize ${#1}$};
\end{tikzpicture}
}

\newcommand{\Charge}{
\begin{tikzpicture}[baseline={([yshift=-.8ex]current bounding box.center)},scale=1.]
\draw [] (0, 0) -- (0, 1.5);
\draw [red] (0, .4) -- (-.5, .4);
\draw [red] (0, 1.1) -- (.5, 1.1);

\node [centered, rotate=0.] at (.2,.15) {\scriptsize $x_i$};
\node [centered, rotate=0.] at (.15,.75) {\scriptsize $y$};
\node [centered, rotate=0.] at (.2,1.35) {\scriptsize $z_j$};
\node [centered, rotate=0.] at (-.6,.4) {\scriptsize $a$};
\node [centered, rotate=0.] at (.6,1.1) {\scriptsize $b$};
\end{tikzpicture}
}

\newcommand{\TrivialCharge}{
\begin{tikzpicture}[baseline={([yshift=-.8ex]current bounding box.center)},scale=1.]
\draw [] (0, 0) -- (0, 1.5);
\draw [red] (0, .4) -- (-.5, .4);
\draw [red] (0, 1.1) -- (.5, 1.1);

\node [centered, rotate=0.] at (.15,.15) {\scriptsize $x$};
\node [centered, rotate=0.] at (.15,.75) {\scriptsize $x$};
\node [centered, rotate=0.] at (.15,1.35) {\scriptsize $x$};
\node [centered, rotate=0.] at (-.6,.4) {\scriptsize $1$};
\node [centered, rotate=0.] at (.6,1.1) {\scriptsize $1$};
\end{tikzpicture}
}

\newcommand{\FibonacciFlux}{
\begin{tikzpicture}[baseline={([yshift=-.8ex]current bounding box.center)},scale=2.]
\draw [ultra thick, ->] (0, 0) -- (1, 0);
\draw [blue, ->] (0, 0) -- (0, 1.);
\node [centered, rotate=0.] at (1.,.1) {\scriptsize $1$};
\node [centered, rotate=0.] at (.1,1.) {\scriptsize $\tau$};

\draw [orange, thick, ->] (.2, .35) -- (.2, .05);
\draw [orange, thick, ->] (.45, .35) -- (.45, .05);
\draw [orange, thick, ->] (.7, .35) -- (.7, .05);
\draw [orange, thick, ->] (.2, -.35) -- (.2, -.05);
\draw [orange, thick, ->] (.45, -.35) -- (.45, -.05);
\draw [orange, thick, ->] (.7, -.35) -- (.7, -.05);
\node [centered, rotate=0.] at (.47,.5) {\scriptsize $P_E^{\Sub_0|\Fibo}$};
\draw [white] (.5, -.4) -- (.6, -.4);
\draw [blue] (.8, 1.1) -- (.8, .9) -- (1, .9);
\node [centered, rotate=0.] at (.93,1.05) {\scriptsize $\tau\bar\tau$};
\end{tikzpicture}
}

\newcommand{\FibonacciDyon}{
\begin{tikzpicture}[baseline={([yshift=-.8ex]current bounding box.center)},scale=2.]
\draw [ultra thick, ->] (0, 0) -- (.4, .9);
\draw [blue, ->] (0, 0) -- (1., 0);
\draw [blue, ->] (0, 0) -- (0, 1.);
\node [centered, rotate=0.] at (1.,.1) {\scriptsize $1$};
\node [centered, rotate=0.] at (.1,1.) {\scriptsize $\tau$};

\draw [orange, thick, ->] (-.175, .325) -- (.05, .225);
\draw [orange, thick, ->] (-.075, .55) -- (.15, .45);
\draw [orange, thick, ->] (.025, .775) -- (.25, .675);
\draw [orange, thick, ->] (.355, .095) -- (.13, .195);
\draw [orange, thick, ->] (.455, .32) -- (.23, .42);
\draw [orange, thick, ->] (.555, .545) -- (.33, .645);
\node [centered, rotate=0.] at (-.25,.65) {\scriptsize $P_E^{\Sub|\Fibo}$};
\draw [blue] (.8, 1.1) -- (.8, .9) -- (1, .9);
\node [centered, rotate=0.] at (.93,1.05) {\scriptsize $\tau\bar\tau$};

\draw [white] (.5, -.4) -- (.6, -.4);
\end{tikzpicture}
}

\newcommand{\PachnerOne}{\begin{tikzpicture}[baseline={([yshift=-.8ex]current bounding box.center)},scale=.5]
\draw [->-] (0.,0.2) -- (.5,1.);
\draw [->-] (0.,1.8) -- (.5,1.);
\draw [->-] (2.,1.) -- (.5,1.);
\draw [->-] (2.5,1.8) -- (2.,1.);
\draw [->-] (2.5,0.2) -- (2.,1.);
\node at (1.25, 1.3) {\scriptsize $m$};
\node at (-.15, 2.) {\scriptsize $a$};
\node at (-.15, 0.) {\scriptsize $b$};
\node at (2.65, 2.) {\scriptsize $c$};
\node at (2.65, 0.) {\scriptsize $d$};
\end{tikzpicture}\qquad = \qquad \sum_{n\in L_{\Fus}}\ \sqrt{d_md_n}\ G^{ijm}_{kln}\quad \begin{tikzpicture}[baseline={([yshift=-.8ex]current bounding box.center)},scale=.5]
\draw [->-] (0.,0.) -- (.8,.5);
\draw [->-] (1.6,0.) -- (.8,.5);
\draw [->-] (.8,.5) -- (.8,2.);
\draw [->-] (0.,2.5) -- (.8,2.);
\draw [->-] (1.6,2.5) -- (.8,2.);
\node at (1.05, 1.3) {\scriptsize $n$};
\node at (-.15, 2.6) {\scriptsize $a$};
\node at (-.15, -.1) {\scriptsize $b$};
\node at (1.75, 2.6) {\scriptsize $c$};
\node at (1.75, -0.1) {\scriptsize $d$};
\end{tikzpicture}}

\newcommand{\PachnerTwo}{\begin{tikzpicture}[baseline={([yshift=-.8ex]current bounding box.center)},scale=.5]
\draw [->-] (1, 0) -- (1, 1);
\draw [->-] (1, 1) arc (270:90:.5);
\draw [->-] (1, 1) arc (-90:90:.5);
\draw [->-] (1, 2) -- (1, 3);
\node at (1.25, .3) {\scriptsize $a$};
\node at (1., 1.5) {\scriptsize \color{red}$\times$};
\node at (.2, 1.5) {\scriptsize $x$};
\node at (1.8, 1.5) {\scriptsize $y$};
\node at (1.25, 2.7) {\scriptsize $b$};
\end{tikzpicture}\qquad = \qquad \sqrt{\frac{d_xd_y}{d_i}}\ \delta_{ij}\ \delta_{xyi^\ast}\quad \begin{tikzpicture}[baseline={([yshift=-.8ex]current bounding box.center)},scale=.5]
\draw [->-] (1, 0) -- (1, 3);
\node at (1.3, 1.5) {\scriptsize $a$};
\end{tikzpicture}}

\newcommand{\PachnerThree}{\begin{tikzpicture}[baseline={([yshift=-.8ex]current bounding box.center)},scale=.5]
\draw [->-] (1, 0) -- (1, 3);
\node at (1.25, 1.5) {\scriptsize $a$};
\end{tikzpicture}\quad = \quad \frac{1}{D}\sum_{xy\in L_{\Fus}} \sqrt{\frac{d_xd_y}{d_i}}\ \delta_{xyi^\ast} \quad \begin{tikzpicture}[baseline={([yshift=-.8ex]current bounding box.center)},scale=.5]
\draw [->-] (1, 0) -- (1, 1);
\draw [->-] (1, 1) arc (270:90:.5);
\draw [->-] (1, 1) arc (-90:90:.5);
\draw [->-] (1, 2) -- (1, 3);
\node at (1.25, .3) {\scriptsize $a$};
\node at (.2, 1.5) {\scriptsize $x$};
\node at (1.8, 1.5) {\scriptsize $y$};
\node at (1.25, 2.7) {\scriptsize $a$};
\end{tikzpicture}}

\newcommand{\PachnerFour}{
\begin{tikzpicture}[baseline={([yshift=-.8ex]current bounding box.center)},scale=.5]
\draw [->-] (3.2,5.6) -- (3.2,6.8);
\draw [->-] (3.2,7.2) -- (4.6,8.);
\draw [->-] (4.6,8.) -- (6.,7.2);
\draw [->-] (6.,5.6) -- (4.6,4.8);
\draw [->-] (4.6,4.8) -- (3.2,5.6);
\draw [->-] (6.,7.2) -- (6.,5.6);
\draw [decorate,decoration={snake,segment length=.12cm, amplitude=.04cm}, ->] (3.2,6.6) -- (4.4,6.6);
\draw [->-] (2.6,7.6) -- (3.2,7.2);
\draw [->-] (4.6,8.8) -- (4.6,8.);
\draw [->-] (6.6,7.6) -- (6.,7.2);
\draw [->-] (6.6,5.2) -- (6.,5.6);
\draw [->-] (4.6,4.) -- (4.6,4.8);
\draw [->-] (2.6,5.2) -- (3.2,5.6);
\draw [->-] (3.2,6.6) -- (3.2,7.2);
\node [centered, rotate=0.] at (2.5,7.8) {\scriptsize $e_1$};
\node [centered, rotate=0.] at (2.9,6.) {\scriptsize $i_0$};
\node [centered, rotate=0.] at (2.9,6.9) {\scriptsize $i_1$};
\node [centered, rotate=0.] at (3.6,7.8) {\scriptsize $i_2$};
\node [centered, rotate=0.] at (4.7,6.6) {\scriptsize $p$};
\end{tikzpicture}\ =\quad\sum_{j\in L_\Fus}\sqrt{d_{i_1}d_j}\ G^{i_2^\ast e_1i_1}_{i_0 p^\ast j}\ 
\begin{tikzpicture}[baseline={([yshift=-.8ex]current bounding box.center)},scale=.5]
\draw [->-] (3.2,5.6) -- (3.2,7.2);
\draw [->-] (3.2,7.2) -- (3.55,7.4);
\draw [->-] (3.55,7.4) -- (4.6,8.);
\draw [->-] (4.6,8.) -- (6.,7.2);
\draw [->-] (6.,5.6) -- (4.6,4.8);
\draw [->-] (4.6,4.8) -- (3.2,5.6);
\draw [->-] (6.,7.2) -- (6.,5.6);
\draw [->-] (2.6,7.6) -- (3.2,7.2);
\draw [->-] (4.6,8.8) -- (4.6,8.);
\draw [->-] (6.6,7.6) -- (6.,7.2);
\draw [->-] (6.6,5.2) -- (6.,5.6);
\draw [->-] (4.6,4.) -- (4.6,4.8);
\draw [->-] (2.6,5.2) -- (3.2,5.6);
\draw [decorate,decoration={snake,segment length=.12cm, amplitude=.04cm}, ->] (3.59,7.42) -- (4.1,6.5);
\node [centered, rotate=0.] at (2.5,5.) {\scriptsize $e_n$};
\node [centered, rotate=0.] at (2.5,7.8) {\scriptsize $e_1$};
\node [centered, rotate=0.] at (2.9,6.5) {\scriptsize $i_0$};
\node [centered, rotate=0.] at (3.3,7.6) {\scriptsize $j$};
\node [centered, rotate=0.] at (3.9,8.) {\scriptsize $i_2$};
\node [centered, rotate=0.] at (4.3,6.3) {\scriptsize $p$};
\node [centered, rotate=0.] at (3.7,5.1) {\scriptsize $i_n$};
\end{tikzpicture}
}

\newcommand{\PachnerFive}{
\begin{tikzpicture}[baseline={([yshift=-.8ex]current bounding box.center)},scale=.5]
\draw [->-] (3.2,5.6) -- (3.2,6.1);
\draw [->-] (3.2,6.1) -- (3.2,6.7);
\draw [->-] (3.2,6.7) -- (3.2,7.2);
\draw [->-] (3.2,7.2) -- (4.6,8.);
\draw [->-] (4.6,8.) -- (6.,7.2);
\draw [->-] (6.,5.6) -- (4.6,4.8);
\draw [->-] (4.6,4.8) -- (3.2,5.6);
\draw [->-] (6.,7.2) -- (6.,5.6);
\draw [decorate,decoration={snake,segment length=.12cm, amplitude=.04cm}, ->] (3.2,6.1) -- (4.4,6.1);
\draw [decorate,decoration={snake,segment length=.12cm, amplitude=.04cm}, ->] (3.2,6.7) -- (4.4,6.7);
\draw [->-] (2.6,7.6) -- (3.2,7.2);
\draw [->-] (4.6,8.8) -- (4.6,8.);
\draw [->-] (6.6,7.6) -- (6.,7.2);
\draw [->-] (6.6,5.2) -- (6.,5.6);
\draw [->-] (4.6,4.) -- (4.6,4.8);
\draw [->-] (2.6,5.2) -- (3.2,5.6);
\node [centered, rotate=0.] at (2.9,5.8) {\scriptsize $i_0$};
\node [centered, rotate=0.] at (2.9,7.1) {\scriptsize $i_1$};
\node [centered, rotate=0.] at (4.7,6.7) {\scriptsize $r$};
\node [centered, rotate=0.] at (4.7,6.1) {\scriptsize $s$};
\node [centered, rotate=0.] at (2.9,6.4) {\scriptsize $j$};
\end{tikzpicture}\quad = \quad \sum_{u\in L_\Fus}\sqrt{d_jd_p}\ G^{r^\ast i_1^\ast j}_{i_0s^\ast p}\quad 
\begin{tikzpicture}[baseline={([yshift=-.8ex]current bounding box.center)},scale=.5]
\draw [->-] (3.2,5.6) -- (3.2,6.8);
\draw [->-] (3.2,7.2) -- (4.6,8.);
\draw [->-] (4.6,8.) -- (6.,7.2);
\draw [->-] (6.,5.6) -- (4.6,4.8);
\draw [->-] (4.6,4.8) -- (3.2,5.6);
\draw [->-] (6.,7.2) -- (6.,5.6);
\draw [decorate,decoration={snake,segment length=.12cm, amplitude=.04cm}, ->] (3.2,6.6) -- (4.4,6.6);
\draw [->-] (2.6,7.6) -- (3.2,7.2);
\draw [->-] (4.6,8.8) -- (4.6,8.);
\draw [->-] (6.6,7.6) -- (6.,7.2);
\draw [->-] (6.6,5.2) -- (6.,5.6);
\draw [->-] (4.6,4.) -- (4.6,4.8);
\draw [->-] (2.6,5.2) -- (3.2,5.6);
\draw [->-] (3.2,6.6) -- (3.2,7.2);
\node [centered, rotate=0.] at (2.9,6.) {\scriptsize $i_0$};
\node [centered, rotate=0.] at (2.9,6.9) {\scriptsize $i_1$};
\node [centered, rotate=0.] at (4.7,6.6) {\scriptsize $p$};
\end{tikzpicture}
}

\newcommand{\PlaquetteMsr}{
\begin{tikzpicture}[baseline={([yshift=-.8ex]current bounding box.center)},scale=.55]
\draw [->-] (3.2,5.6) -- (3.2,6.8);
\draw [->-] (3.2,7.2) -- (4.6,8.);
\draw [->-] (4.6,8.) -- (6.,7.2);
\draw [->-] (6.,5.6) -- (4.6,4.8);
\draw [->-] (4.6,4.8) -- (3.2,5.6);
\draw [->-] (6.,7.2) -- (6.,5.6);
\draw [->-] (2.6,7.6) -- (3.2,7.2);
\draw [->-] (4.6,8.8) -- (4.6,8.);
\draw [->-] (6.6,7.6) -- (6.,7.2);
\draw [->-] (6.6,5.2) -- (6.,5.6);
\draw [->-] (4.6,4.) -- (4.6,4.8);
\draw [->-] (2.6,5.2) -- (3.2,5.6);
\draw [->-] (3.2,6.6) -- (3.2,7.2);
\node [centered, rotate=0.] at (2.4,7.8) {\scriptsize $e_1$};
\node [centered, rotate=0.] at (4.95,8.8) {\scriptsize $e_2$};
\node [centered, rotate=0.] at (6.9,7.8) {\scriptsize $e_3$};
\node [centered, rotate=0.] at (6.9,5.) {\scriptsize $e_4$};
\node [centered, rotate=0.] at (5.,4.) {\scriptsize $e_5$};
\node [centered, rotate=0.] at (2.3,5.) {\scriptsize $e_6$};
\node [centered, rotate=0.] at (2.85,6.) {\scriptsize $i_7$};
\node [centered, rotate=0.] at (2.85,6.9) {\scriptsize $i_1$};
\node [centered, rotate=0.] at (3.6,7.9) {\scriptsize $i_2$};
\node [centered, rotate=0.] at (5.6,7.9) {\scriptsize $i_3$};
\node [centered, rotate=0.] at (6.4,6.4) {\scriptsize $i_4$};
\node [centered, rotate=0.] at (5.6,4.9) {\scriptsize $i_5$};
\node [centered, rotate=0.] at (3.7,4.9) {\scriptsize $i_6$};

\draw [decorate,decoration={snake,segment length=.12cm, amplitude=.04cm}, ->] (3.2,6.4) -- (5.,6.4);
\draw [->] (3.8, 6.4) arc (180:0:.8);
\draw [] (4.4, 6.4) arc (-180:0:0.5);
\node [centered, rotate=0.] at (4.8,6.7) {\scriptsize $p$};
\node [centered, rotate=0.] at (4.1,6.1) {\scriptsize $t$};
\node [centered, rotate=0.] at (3.5,6.1) {\scriptsize $p$};
\node [centered, rotate=0.] at (5.65,6.4) {\scriptsize $s$};
\node [centered, rotate=0.] at (4.6,5.4) {\color{red} $\times$};
\end{tikzpicture}
}

\newcommand{\ExcitedA}{
\begin{tikzpicture}[baseline={([yshift=-.8ex]current bounding box.center)},scale=.8]
\draw [->-] (3.2,5.6) -- (3.2,7.2);
\draw [->-] (3.2,7.2) -- (3.9,7.6);
\draw [->-] (2.5,7.6) -- (3.2,7.2);
\draw [->-] (2.5,5.2) -- (3.2,5.6);
\draw [->-] (3.2,5.6) -- (3.9,5.2);
\node [centered, rotate=0.] at (2.4,5.) {\scriptsize $e_n$};
\node [centered, rotate=0.] at (4.1,5.) {\scriptsize $i_n$};
\node [centered, rotate=0.] at (4.1,7.8) {\scriptsize $i_2$};
\node [centered, rotate=0.] at (2.4,7.8) {\scriptsize $e_1$};
\node [centered, rotate=0.] at (2.9,6.4) {\scriptsize $j$};
\end{tikzpicture}
}

\newcommand{\ExcitedB}{
\begin{tikzpicture}[baseline={([yshift=-.8ex]current bounding box.center)},scale=.8]
\draw [->-] (3.2,5.6) -- (3.2,6.1);
\draw [->-] (3.2,6.1) -- (3.2,6.7);
\draw [->-] (3.2,6.7) -- (3.2,7.2);
\draw [->-] (3.2,7.2) -- (3.9,7.6);
\draw [->-] (2.5,7.6) -- (3.2,7.2);
\draw [->-] (2.5,5.2) -- (3.2,5.6);
\draw [->-] (3.2,5.6) -- (3.9,5.2);
\draw [decorate,decoration={snake,segment length=.12cm, amplitude=.04cm}, ->] (3.2,6.7) -- (4.,6.7);
\draw [decorate,decoration={snake,segment length=.12cm, amplitude=.04cm}, ->] (3.2,6.1) -- (2.4,6.1);
\node [centered, rotate=0.] at (2.4,5.) {\scriptsize $e_n$};
\node [centered, rotate=0.] at (4.1,5.) {\scriptsize $i_n$};
\node [centered, rotate=0.] at (4.1,7.8) {\scriptsize $i_2$};
\node [centered, rotate=0.] at (2.4,7.8) {\scriptsize $e_1$};
\node [centered, rotate=0.] at (3.,5.8) {\scriptsize $j$};
\node [centered, rotate=0.] at (3.,6.4) {\scriptsize $k$};
\node [centered, rotate=0.] at (3.4,7.) {\scriptsize $j$};
\node [centered, rotate=0.] at (4.2,6.7) {\scriptsize $q$};
\node [centered, rotate=0.] at (2.2,6.1) {\scriptsize $p^\ast$};
\end{tikzpicture}
}

\newcommand{\FrobeniusA}{\begin{tikzpicture}[baseline={([yshift=-.8ex]current bounding box.center)},scale=.8]
\draw [red, ->-] (.3, 0.5) -- (1, 1);
\draw [red, ->-] (1.7, 0.5) -- (1, 1);
\draw [red, ->-] (1., 2.) -- (1, 1);
\fill [red] (1., 1.) circle (.15);
\node at (1.3, 1.7) {\scriptsize $a_\alpha$};
\node at (.2, .8) {\scriptsize $b_\beta$};
\node at (1.75, .8) {\scriptsize $c_\gamma$};
\end{tikzpicture}\qquad := \qquad f_{a_\alpha b_\beta c_\gamma}\quad \begin{tikzpicture}[baseline={([yshift=-.8ex]current bounding box.center)},scale=.8]
\draw [->-] (.3, 0.5) -- (1, 1);
\draw [->-] (1.7, 0.5) -- (1, 1);
\draw [->-] (1., 2.) -- (1, 1);
\node at (1.25, 1.6) {\scriptsize $a_\alpha$};
\node at (.3, .8) {\scriptsize $b_\beta$};
\node at (1.6, .8) {\scriptsize $c_\gamma$};
\end{tikzpicture}}

\newcommand{\FrobeniusB}{\T\ \sum_{\gamma = 0}^{n_c}\begin{tikzpicture}[baseline={([yshift=-.8ex]current bounding box.center)},scale=.7]
\draw [red, ->-] (0.,0.2) -- (.5,1.);
\draw [red, ->-] (0.,1.8) -- (.5,1.);
\draw [red, ->-] (2.,1.) -- (.5,1.);
\draw [red, ->-] (2.5,1.8) -- (2.,1.);
\draw [red, ->-] (2.5,0.2) -- (2.,1.);
\fill [red] (.5, 1.) circle (.15);
\fill [red] (2., 1.) circle (.15);
\node at (1.25, 1.3) {\scriptsize $c_\gamma$};
\node at (-.1, 2.) {\scriptsize $s_\sigma$};
\node at (-.1, 0.) {\scriptsize $a_\alpha$};
\node at (2.6, 2.) {\scriptsize $r_\rho$};
\node at (2.8, 0.) {\scriptsize $b_\beta$};
\end{tikzpicture}\qquad =\qquad \sum_{t_\tau\in L_{\mathcal{A}}}\quad \begin{tikzpicture}[baseline={([yshift=-.8ex]current bounding box.center)},scale=.7]
\draw [red, ->-] (0.,0.) -- (.8,.5);
\draw [red, ->-] (1.6,0.) -- (.8,.5);
\draw [red, ->-] (.8,.5) -- (.8,2.);
\draw [red, ->-] (0.,2.5) -- (.8,2.);
\draw [red, ->-] (1.6,2.5) -- (.8,2.);
\fill [red] (.8, .5) circle (.15);
\fill [red] (.8, 2.) circle (.15);
\node at (1.2, 1.3) {\scriptsize $t_\tau$};
\node at (-.2, 2.7) {\scriptsize $s_\sigma$};
\node at (-.2, -.2) {\scriptsize $a_\alpha$};
\node at (1.9, 2.7) {\scriptsize $r_\rho$};
\node at (1.9, -0.2) {\scriptsize $b_\beta$};
\end{tikzpicture}\qquad =: \qquad \begin{tikzpicture}[baseline={([yshift=-.8ex]current bounding box.center)},scale=.7]
\draw [red, ->-] (0.,0.) -- (.8,.5);
\draw [red, ->-] (1.6,0.) -- (.8,.5);
\draw [red, thick, densely dashed] (.8,.5) -- (.8,2.);
\draw [red, ->-] (0.,2.5) -- (.8,2.);
\draw [red, ->-] (1.6,2.5) -- (.8,2.);
\fill [red] (.8, .5) circle (.15);
\fill [red] (.8, 2.) circle (.15);
\node at (-.2, 2.7) {\scriptsize $s_\sigma$};
\node at (-.2, -.2) {\scriptsize $a_\alpha$};
\node at (1.9, 2.7) {\scriptsize $r_\rho$};
\node at (1.9, -0.2) {\scriptsize $b_\beta$};
\end{tikzpicture}}

\newcommand{\FrobeniusC}{\T\quad \begin{tikzpicture}[baseline={([yshift=-.8ex]current bounding box.center)},scale=.7]
\draw [red, ->-] (1, 0) -- (1, 1);
\draw [red, thick, densely dashed] (1, 1) arc (-90:270:.5);
\draw [red, ->-] (1, 2) -- (1, 3);
\fill [red] (1., 1.) circle (.1);
\fill [red] (1., 2.) circle (.1);
\node at (1.35, .3) {\scriptsize $a_\alpha$};
\node at (1.35, 2.7) {\scriptsize $b_\beta$};
\node [red] at (1., 1.5) {$\times$};
\end{tikzpicture}\qquad = \qquad d_{\mathcal{A}}\ \delta_{ab}\ \delta_{\alpha\beta}\quad \begin{tikzpicture}[baseline={([yshift=-.8ex]current bounding box.center)},scale=.7]
\draw [red, ->-] (1, 0) -- (1, 3);
\node at (1.35, 1.5) {\scriptsize $a_\alpha$};
\end{tikzpicture}}

\newcommand{\BimoduleA}{\begin{tikzpicture}[baseline={([yshift=-.8ex]current bounding box.center)},scale=.8]
\draw [blue, ->-] (1, 0) -- (1, .6);
\draw [blue, snake it, ultra thick] (1, .6) -- (1, 1.3);
\draw [blue, ->-] (1, 1.3) -- (1, 2.);
\draw [red, -<-] (1, .6) -- (.3, .6);
\draw [red, -<-] (1, 1.3) -- (1.7, 1.3);
\node at (0.1, .6) {\scriptsize $a_\alpha$};
\node at (1.9, 1.3) {\scriptsize $b_\beta$};
\node at (1.3, .2) {\scriptsize $x_\chi$};
\node at (1.25,.9) {\scriptsize $M$};
\node at (1.25,1.8) {\scriptsize $z_\zeta$};
\end{tikzpicture}\qquad := \qquad\sum_{y\in L_\Fus}\ [P_M]^{a_\alpha b_\beta}_{x_\chi y z_\zeta}\quad\begin{tikzpicture}[baseline={([yshift=-.8ex]current bounding box.center)},scale=.8]
\draw [->-] (1, 0) -- (1, .6);
\draw [->-] (1, .6) -- (1, 1.3);
\draw [->-] (1, 1.3) -- (1, 2.);
\draw [-<-] (1, .6) -- (.3, .6);
\draw [-<-] (1, 1.3) -- (1.7, 1.3);
\node at (0.1, .6) {\scriptsize $a_\alpha$};
\node at (1.9, 1.3) {\scriptsize $b_\beta$};
\node at (1.25, .2) {\scriptsize $x_\chi$};
\node at (1.15,.9) {\scriptsize $y$};
\node at (1.25,1.8) {\scriptsize $z_\zeta$};
\end{tikzpicture}
}

\newcommand{\BimoduleB}{\T\ \sum_{y_\upsilon\in L_M}\quad
\begin{tikzpicture}[baseline={([yshift=-.8ex]current bounding box.center)},scale=.6]
\draw [blue, ->-] (1, 0) -- (1, 1.);
\draw [blue, snake it, ultra thick] (1, 1.) -- (1, 1.8);
\draw [blue, ->-] (1, 1.8) -- (1, 2.6);
\draw [blue, snake it, ultra thick] (1, 2.6) -- (1, 3.4);
\draw [blue, ->-] (1, 3.4) -- (1, 4.4);

\draw [red, -<-] (1, 1.) -- (0, 1.);
\draw [red, -<-] (1, 1.8) -- (2, 1.8);
\draw [red, -<-] (1, 2.6) -- (0, 2.6);
\draw [red, -<-] (1, 3.4) -- (2, 3.4);

\node at (1.4, .5) {\scriptsize $x_\chi$};
\node at (-.4, 1.) {\scriptsize $a_\alpha$};
\node at (1.4, 1.4) {\scriptsize $M$};
\node at (2.4, 1.8) {\scriptsize $r_\rho$};
\node at (1.4, 2.2) {\scriptsize $y_\upsilon$};
\node at (-.4, 2.6) {\scriptsize $b_\beta$};
\node at (1.4, 3.) {\scriptsize $M$};
\node at (2.4, 3.4) {\scriptsize $s_\sigma$};
\node at (1.4, 3.9) {\scriptsize $z_\zeta$};
\end{tikzpicture}\qquad = \qquad
\begin{tikzpicture}[baseline={([yshift=-.8ex]current bounding box.center)},scale=.6]
\draw [blue, ->-] (1, 0) -- (1, 1.5);
\draw [blue, snake it, ultra thick] (1, 1.5) -- (1, 2.9);
\draw [blue, ->-] (1, 2.9) -- (1, 4.4);
\draw [red, -<-] (-.5, 1.5) -- (-1.3, 2.5);
\draw [red, -<-] (-.5, 1.5) -- (-1.3, .5);
\draw [red, densely dashed, very thick] (1, 1.5) -- (-.5, 1.5);
\fill [red] (-.5, 1.5) circle (0.15);
\draw [red, densely dashed, very thick] (1, 2.9) -- (2.5, 2.9);
\draw [red, -<-] (2.5, 2.9) -- (3.3, 3.9);
\draw [red, -<-] (2.5, 2.9) -- (3.3, 1.9);
\fill [red] (2.5, 2.9) circle (0.15);
\node at (1.4, .7) {\scriptsize $x_\chi$};
\node at (1.4, 2.2) {\scriptsize $M$};
\node at (1.4, 3.7) {\scriptsize $z_\zeta$};
\node at (-1.6, .3) {\scriptsize $a_\alpha$};
\node at (-1.6, 2.7) {\scriptsize $b_\beta$};
\node at (3.6, 1.7) {\scriptsize $r_\rho$};
\node at (3.6, 4.1) {\scriptsize $s_\sigma$};
\end{tikzpicture}}

\newcommand{\BimoduleF}{\T\sum_{\alpha = 1}^{n^M_x}\ \sum_{\beta = 1}^{n^N_x}\quad
\begin{tikzpicture}[baseline={([yshift=-.8ex]current bounding box.center)},scale=1.]
\draw [blue, ->-] (0, 0) -- (0, .5);
\draw [blue, ultra thick, snake it] (0, .5) -- (0, 1.);
\draw [blue, ->-] (0, 1) -- (0, 1.75);
\draw [blue, ->-] (0, 1.75) -- (0, 2.5);
\draw [blue, ultra thick, snake it] (0, 2.5) -- (0, 3);
\draw [blue, ->-] (0, 3.) -- (0, 3.5);

\draw [red, thick, densely dashed] (0, 1) arc (-90:90:1.);
\draw [red, thick, densely dashed] (0, .5) arc (270:90:1.);
\draw [red, ->-] (-1.7, 1.5) -- (-1., 1.5);
\draw [red, ->-] (1.7, 2.) -- (1., 2.);
\node [] at (-1.5, 1.7) {$a$};
\node [] at (1.5, 2.2) {$b$};
\node [] at (.3, .2) {$y^M_\upsilon$};
\node [] at (.3, .75) {$M$};
\node [] at (-.3, 1.4) {$x^M_\alpha$};
\node [] at (.3, 2.1) {$x^N_\beta$};
\node [] at (-.3, 2.75) {$N$};
\node [] at (.3, 3.4) {$z^N_\zeta$};
\end{tikzpicture}\quad = \quad \delta_{MN}\frac{d_\mathcal{A}^2n^M_xd_x}{d_M}\quad
\begin{tikzpicture}[baseline={([yshift=-.8ex]current bounding box.center)},scale=1.]
\draw [blue, ->-] (0, 0) -- (0, 1.);
\draw [blue, ultra thick, snake it] (0, 1.) -- (0, 2.5);
\draw [blue, ->-] (0, 2.5) -- (0, 3.5);
\draw [red, ->-] (-1.5, 1.) -- (0., 1.);
\draw [red, ->-] (1.5, 2.5) -- (0., 2.5);
\node [] at (-1.5, 1.2) {$a$};
\node [] at (1.5, 2.7) {$b$};
\node [] at (.4, .5) {$y^M_\upsilon$};
\node [] at (.3, 1.75) {$M$};
\node [] at (.4, 3.) {$z^M_\zeta$};
\end{tikzpicture}
}

\newcommand{\BimoduleG}{
\begin{tikzpicture}[baseline={([yshift=-.8ex]current bounding box.center)},scale=1.]
\draw [blue, ->-] (0, 0) -- (0, 0.5);
\draw [blue, snake it, ultra thick] (0, 0.5) -- (0, 1.);
\draw [blue, ->-] (0, 1) -- (0, 1.5);
\draw [red, -<-] (0, .5) -- (-.5, .5);
\draw [red, -<-] (0, 1.) -- (.5, 1.);

\node [centered, rotate=0.] at (.2,.15) {\scriptsize $x_\chi$};
\node [centered, rotate=0.] at (.2,.75) {\scriptsize $M$};
\node [centered, rotate=0.] at (.2,1.35) {\scriptsize $z_\zeta$};
\node [centered, rotate=0.] at (-.7,.5) {\scriptsize $a_\alpha$};
\node [centered, rotate=0.] at (.7,1.) {\scriptsize $b_\beta$};
\end{tikzpicture}
}

\newcommand{\BimoduleH}{\D\qquad
\begin{tikzpicture}[baseline={([yshift=-.8ex]current bounding box.center)},scale=.5]
\draw [->-] (3.2,5.6) -- (3.2,6.5);
\draw [->-] (3.2,6.5) -- (3.2,7.2);
\draw [->-] (3.2,7.2) -- (4.6,8.);
\draw [->-] (4.6,8.) -- (6.,7.2);
\draw [->-] (6.,7.2) -- (6.,5.6);
\draw [->-] (6.,5.6) -- (4.6,4.8);
\draw [->-] (4.6,4.8) -- (3.2,5.6);
\draw [decorate,decoration={snake,segment length=.12cm, amplitude=.04cm}, ->] (3.2,6.4) -- (4.2,6.4);
\node [centered, rotate=0.] at (2.7,6.) {\scriptsize $I_6$};
\node [centered, rotate=0.] at (2.7,6.9) {\scriptsize $I_0$};
\node [centered, rotate=0.] at (3.6,7.9) {\scriptsize $I_1$};
\node [centered, rotate=0.] at (5.6,7.9) {\scriptsize $I_2$};
\node [centered, rotate=0.] at (6.4,6.4) {\scriptsize $I_3$};
\node [centered, rotate=0.] at (5.6,4.9) {\scriptsize $I_4$};
\node [centered, rotate=0.] at (3.6,4.9) {\scriptsize $I_5$};
\node [centered, rotate=0.] at (4.5,6.4) {\scriptsize $M$};
\draw [->-] (2.6,7.6) -- (3.2,7.2);
\draw [->-] (4.6,8.8) -- (4.6,8.);
\draw [->-] (6.6,7.6) -- (6.,7.2);
\draw [->-] (6.6,5.2) -- (6.,5.6);
\draw [->-] (4.6,4.) -- (4.6,4.8);
\draw [->-] (2.6,5.2) -- (3.2,5.6);
\node [centered, rotate=0.] at (2.3,7.8) {\scriptsize $E_1$};
\node [centered, rotate=0.] at (4.6,9.) {\scriptsize $E_2$};
\node [centered, rotate=0.] at (6.9,7.8) {\scriptsize $E_3$};
\node [centered, rotate=0.] at (6.9,5.) {\scriptsize $E_4$};
\node [centered, rotate=0.] at (4.6,3.7) {\scriptsize $E_5$};
\node [centered, rotate=0.] at (2.3,5.) {\scriptsize $E_6$};
\end{tikzpicture}\qquad = \qquad\frac{1}{d_\mathcal{A}^{9}}\sum_{x_i, y_i\in L_{M_i}}\sum_{e_i\in L_{N_i}}\sum_{p_\alpha,q_\beta\in L_M}\nonumber\\
\T\qquad
\begin{tikzpicture}[baseline={([yshift=-.8ex]current bounding box.center)},scale=1.4]
\draw [blue, ->-] (3.2,5.6) -- (3.2,5.8);
\draw [blue, ultra thick, snake it] (3.2,5.8) -- (3.2,6.1);
\draw [blue, ->-] (3.2,6.1) -- (3.2,6.4);
\node [centered, rotate=0.] at (3.05,5.7) {\scriptsize $x_6$};
\node [centered, rotate=0.] at (3.02,5.95) {\scriptsize $I_6$};
\node [centered, rotate=0.] at (3.05,6.25) {\scriptsize $y_6$};
\draw [red, thick, densely dashed] (3.2, 5.8) -- (2.65, 5.8);
\draw [red, thick, densely dashed] (3.2, 6.1) -- (4.1, 6.1);
\fill [red] (4.05, 6.1) circle (0.05);
\draw [blue, ->-] (3.2,6.4) -- (3.2,6.7);
\draw [blue, ultra thick, snake it] (3.2,6.7) -- (3.2,7.);
\draw [blue, ->-] (3.2,7.) -- (3.2,7.2);
\node [centered, rotate=0.] at (3.05,6.55) {\scriptsize $x_0$};
\node [centered, rotate=0.] at (3.02,6.85) {\scriptsize $I_0$};
\node [centered, rotate=0.] at (3.05,7.1) {\scriptsize $y_0$};
\draw [red, thick, densely dashed] (3.2, 6.7) -- (2.65, 6.7);
\draw [red, thick, densely dashed] (3.2, 7.) -- (3.8, 7.);
\fill [red] (3.8, 7.) circle (0.05);
\draw [blue, ->-] (3.2,7.2) -- (3.55,7.4);
\draw [blue, ultra thick, snake it] (3.55,7.4) -- (4.25,7.8);
\draw [blue, ->-] (4.25,7.8) -- (4.6,8.);
\node [centered, rotate=0.] at (3.45,7.2) {\scriptsize $x_1$};
\node [centered, rotate=0.] at (3.8,7.72) {\scriptsize $I_1$};
\node [centered, rotate=0.] at (4.35,8.) {\scriptsize $y_1$};
\draw [red, thick, densely dashed] (3.55, 7.4) -- (3.25, 7.9);
\draw [red, thick, densely dashed] (4.25, 7.8) -- (4.45, 7.45);
\fill [red] (4.47, 7.4) circle (0.05);
\draw [blue, ->-] (4.6,8.) -- (4.95,7.8);
\draw [blue, ultra thick, snake it] (4.95,7.8) -- (5.65,7.4);
\draw [blue, ->-] (5.65,7.4) -- (6.,7.2);
\node [centered, rotate=0.] at (5.9,7.4) {\scriptsize $y_2$};
\node [centered, rotate=0.] at (5.4,7.72) {\scriptsize $I_2$};
\node [centered, rotate=0.] at (4.77,7.8) {\scriptsize $x_2$};
\draw [red, thick, densely dashed] (4.95, 7.8) -- (5.25, 8.3);
\draw [red, thick, densely dashed] (5.65, 7.4) -- (5.38, 7.);
\fill [red] (5.39, 7.02) circle (0.05);
\draw [blue, ->-] (6.,7.2) -- (6.,6.7);
\draw [blue, ultra thick, snake it] (6.,6.7) -- (6.,6.1);
\draw [blue, ->-] (6.,6.1) -- (6.,5.6);
\draw [red, thick, densely dashed] (6., 6.7) -- (6.5, 6.7);
\draw [red, thick, densely dashed] (5.6, 6.1) -- (6., 6.1);
\fill [red] (5.56, 6.1) circle (0.05);
\node [centered, rotate=0.] at (5.85,6.9) {\scriptsize $x_3$};
\node [centered, rotate=0.] at (6.2,6.4) {\scriptsize $I_3$};
\node [centered, rotate=0.] at (6.15,5.85) {\scriptsize $y_3$};
\draw [blue, ->-] (6.,5.6) -- (5.65,5.4);
\draw [blue, ultra thick, snake it] (5.65,5.4) -- (4.95,5.);
\draw [blue, ->-] (4.95,5.) -- (4.6,4.8);
\node [centered, rotate=0.] at (5.95,5.35) {\scriptsize $x_4$};
\node [centered, rotate=0.] at (5.4,5.05) {\scriptsize $I_4$};
\node [centered, rotate=0.] at (4.9,4.78) {\scriptsize $y_4$};
\draw [red, thick, densely dashed] (5.65, 5.4) -- (5.95, 4.9);
\draw [red, thick, densely dashed] (4.95, 5.) -- (4.55, 5.7);
\fill [red] (4.58, 5.64) circle (0.05);
\draw [blue, ->-] (4.6,4.8) -- (4.25,5.);
\draw [blue, ultra thick, snake it] (4.25,5.) -- (3.55,5.4);
\draw [blue, ->-] (3.55,5.4) -- (3.2,5.6);
\node [centered, rotate=0.] at (3.48,5.6) {\scriptsize $y_5$};
\node [centered, rotate=0.] at (3.75,5.05) {\scriptsize $I_5$};
\node [centered, rotate=0.] at (4.5,5.03) {\scriptsize $x_5$};
\draw [red, thick, densely dashed] (4.25, 5.) -- (3.95, 4.5);
\draw [red, thick, densely dashed] (3.55, 5.4) -- (4.25, 5.8);
\fill [red] (4.25, 5.8) circle (0.05);
\draw [blue, ->-] (3.2,6.4) -- (3.6,6.4);
\draw [blue, ultra thick, snake it] (3.6,6.4) -- (4.,6.4);
\draw [blue, ->] (4.,6.4) -- (4.5,6.4);
\draw [red, thick, densely dashed] (3.6, 6.4) arc (180:0:1.);
\draw [red, thick, densely dashed] (4., 6.4) arc (-180:0:.8);
\node [centered, rotate=0.] at (3.4,6.5) {\scriptsize $p_\alpha$};
\node [centered, rotate=0.] at (3.8,6.53) {\scriptsize $M$};
\node [centered, rotate=0.] at (4.6,6.4) {\scriptsize $q_\beta$};
\node [centered, rotate=0.] at (3.4,6.8) {\color{red} $\times$};
\node [centered, rotate=0.] at (3.8,7.3) {\color{red} $\times$};
\node [centered, rotate=0.] at (4.9,7.5) {\color{red} $\times$};
\node [centered, rotate=0.] at (5.75,6.7) {\color{red} $\times$};
\node [centered, rotate=0.] at (5.4,5.6) {\color{red} $\times$};
\node [centered, rotate=0.] at (4.3,5.4) {\color{red} $\times$};
\node [centered, rotate=0.] at (3.6,5.8) {\color{red} $\times$};
\node [centered, rotate=0.] at (3.6,6.2) {\color{red} $\times$};

\draw [blue, ultra thick, snake it] (2.6,7.6) -- (2.9,7.4);
\draw [blue, ->-] (2.9,7.4) -- (3.2,7.2);
\draw [red, thick, densely dashed] (2.9,7.4) -- (2.68,7.07);
\node [centered, rotate=0.] at (2.5,7.7) {\scriptsize $E_1$};
\node [centered, rotate=0.] at (3.1,7.43) {\scriptsize $e_1$};
\draw [blue, ultra thick, snake it] (4.6,8.6) -- (4.6,8.3);
\draw [blue, ->-] (4.6,8.3) -- (4.6,8.);
\draw [red, thick, densely dashed] (4.6,8.3) -- (4.2,8.3);
\node [centered, rotate=0.] at (4.6,8.7) {\scriptsize $E_2$};
\node [centered, rotate=0.] at (4.72,8.1) {\scriptsize $e_2$};
\draw [blue, ultra thick, snake it] (6.6,7.6) -- (6.3,7.4);
\draw [blue, ->-] (6.3,7.4) -- (6.,7.2);
\draw [red, thick, densely dashed] (6.3,7.4) -- (6.08,7.73);
\node [centered, rotate=0.] at (6.7,7.7) {\scriptsize $E_3$};
\node [centered, rotate=0.] at (6.2,7.19) {\scriptsize $e_3$};
\draw [blue, ultra thick, snake it] (6.6,5.2) -- (6.3,5.4);
\draw [blue, ->-] (6.3,5.4) -- (6.,5.6);
\draw [red, thick, densely dashed] (6.3,5.4) -- (6.7,5.73);
\node [centered, rotate=0.] at (6.8,5.1) {\scriptsize $E_4$};
\node [centered, rotate=0.] at (6.3,5.6) {\scriptsize $e_4$};
\draw [blue, ultra thick, snake it] (4.6,4.2) -- (4.6,4.5);
\draw [blue, ->-] (4.6,4.5) -- (4.6,4.8);
\draw [red, thick, densely dashed] (4.6,4.5) -- (5.,4.5);
\node [centered, rotate=0.] at (4.65,4.1) {\scriptsize $E_5$};
\node [centered, rotate=0.] at (4.38,4.67) {\scriptsize $e_5$};
\draw [blue, ultra thick, snake it] (2.6,5.2) -- (2.9,5.4);
\draw [blue, ->-] (2.9,5.4) -- (3.2,5.6);
\draw [red, thick, densely dashed] (2.9,5.4) -- (3.12,5.07);
\node [centered, rotate=0.] at (2.5,5.1) {\scriptsize $E_6$};
\node [centered, rotate=0.] at (3.12,5.38) {\scriptsize $e_6$};
\end{tikzpicture}
}

\newcommand{\FibonacciC}{\begin{tikzpicture}[baseline={([yshift=-.8ex]current bounding box.center)},scale=.5]
\draw [] (3.2,5.6) -- (3.2,6.5);
\draw [] (3.2,6.5) -- (3.2,7.2);
\draw [] (3.2,7.2) -- (4.6,8.);
\draw [] (4.6,8.) -- (6.,7.2);
\draw [] (6.,7.2) -- (6.,5.6);
\draw [] (6.,5.6) -- (4.6,4.8);
\draw [] (4.6,4.8) -- (3.2,5.6);
\draw [decorate,decoration={snake,segment length=.12cm, amplitude=.04cm}] (3.2,6.4) -- (4.4,6.4);
\node [centered, rotate=0.] at (2.7,6.) {\scriptsize $I_6$};
\node [centered, rotate=0.] at (2.7,6.9) {\scriptsize $I_0$};
\node [centered, rotate=0.] at (3.6,7.9) {\scriptsize $I_1$};
\node [centered, rotate=0.] at (5.6,7.9) {\scriptsize $I_2$};
\node [centered, rotate=0.] at (6.5,6.4) {\scriptsize $I_3$};
\node [centered, rotate=0.] at (5.6,4.9) {\scriptsize $I_4$};
\node [centered, rotate=0.] at (3.7,4.9) {\scriptsize $I_5$};
\node [centered, rotate=0.] at (4.7,6.4) {\scriptsize $M$};
\draw [] (2.6,7.6) -- (3.2,7.2);
\draw [] (4.6,8.8) -- (4.6,8.);
\draw [] (6.6,7.6) -- (6.,7.2);
\draw [] (6.6,5.2) -- (6.,5.6);
\draw [] (4.6,4.) -- (4.6,4.8);
\draw [] (2.6,5.2) -- (3.2,5.6);
\node [centered, rotate=0.] at (2.2,7.8) {\scriptsize $E_1$};
\node [centered, rotate=0.] at (4.7,9.) {\scriptsize $E_2$};
\node [centered, rotate=0.] at (7.,7.8) {\scriptsize $E_3$};
\node [centered, rotate=0.] at (7.,5.) {\scriptsize $E_4$};
\node [centered, rotate=0.] at (4.7,3.8) {\scriptsize $E_5$};
\node [centered, rotate=0.] at (2.2,5.) {\scriptsize $E_6$};
\end{tikzpicture}\qquad \Longrightarrow \qquad\sum_{i_k, e_k = 1, \tau}\ \ \sum_{p\in L_M}\cdots\begin{tikzpicture}[baseline={([yshift=-.8ex]current bounding box.center)},scale=.5]
\draw [] (3.2,5.6) -- (3.2,6.5);
\draw [] (3.2,6.5) -- (3.2,7.2);
\draw [] (3.2,7.2) -- (4.6,8.);
\draw [] (4.6,8.) -- (6.,7.2);
\draw [] (6.,7.2) -- (6.,5.6);
\draw [] (6.,5.6) -- (4.6,4.8);
\draw [] (4.6,4.8) -- (3.2,5.6);
\draw [decorate,decoration={snake,segment length=.12cm, amplitude=.04cm}] (3.2,6.4) -- (4.4,6.4);
\node [centered, rotate=0.] at (2.8,6.) {\scriptsize $i_6$};
\node [centered, rotate=0.] at (2.8,6.9) {\scriptsize $i_0$};
\node [centered, rotate=0.] at (3.6,7.9) {\scriptsize $i_1$};
\node [centered, rotate=0.] at (5.6,7.9) {\scriptsize $i_2$};
\node [centered, rotate=0.] at (6.4,6.4) {\scriptsize $i_3$};
\node [centered, rotate=0.] at (5.6,4.9) {\scriptsize $i_4$};
\node [centered, rotate=0.] at (3.7,4.9) {\scriptsize $i_5$};
\node [centered, rotate=0.] at (4.7,6.4) {\scriptsize $p$};
\draw [] (2.6,7.6) -- (3.2,7.2);
\draw [] (4.6,8.8) -- (4.6,8.);
\draw [] (6.6,7.6) -- (6.,7.2);
\draw [] (6.6,5.2) -- (6.,5.6);
\draw [] (4.6,4.) -- (4.6,4.8);
\draw [] (2.6,5.2) -- (3.2,5.6);
\node [centered, rotate=0.] at (2.2,7.8) {\scriptsize $e_1$};
\node [centered, rotate=0.] at (4.7,9.) {\scriptsize $e_2$};
\node [centered, rotate=0.] at (7.,7.8) {\scriptsize $e_3$};
\node [centered, rotate=0.] at (7.,5.) {\scriptsize $e_4$};
\node [centered, rotate=0.] at (4.7,3.8) {\scriptsize $e_5$};
\node [centered, rotate=0.] at (2.2,5.) {\scriptsize $e_6$};
\end{tikzpicture}}

\newcommand{\FigureCorrespondence}{
\begin{tikzpicture}[baseline={([yshift=-.8ex]current bounding box.center)},scale=.95]

\filldraw [red!20!white!100!] (5.6,6.4) -- (9.0,6.4) -- (10.2,7.6) -- (6.8,7.6);
\node [centered, rotate=0.] at (7.5,6.7) {\scriptsize {Parent Phase}};
\fill [red] (9.0, 7.0) circle (0.1);
\node [centered, rotate=0.] at (9.3,7.3) {\scriptsize $J_\text{Parent}$};

\filldraw [black!20!white!100!] (5.6,5.6) -- (9.0,5.6) -- (10.2,6.8) -- (9.4,6.8) -- (9.0,6.4) -- (6.4,6.4);
\node [centered, rotate=0.] at (7.5,5.9) {\scriptsize {Intermediate Phase}};
\fill [black] (9.0, 6.) circle (0.1);
\node [centered, rotate=0.] at (9.3,6.3) {\scriptsize $J_\text{Int}$};

\filldraw [blue!20!white!100!] (5.6,4.8) -- (9.0,4.8) -- (10.2,6.) -- (9.4,6.) -- (9.0,5.6) -- (6.4,5.6);
\node [centered, rotate=0.] at (7.5,5.1) {\scriptsize {Child Phase}};
\fill [blue] (9.0, 5.1) circle (0.1);
\node [centered, rotate=0.] at (9.2,5.4) {\scriptsize $J_\text{Child}$};

\draw [ultra thick, ->, magenta] (9.9,8.) -- (9.9,4.8);
\node [centered, rotate=0.] at (10.1,6.3) {\scriptsize $t$};
\end{tikzpicture}
}

\newcommand{\HalfBraidingA}{
\begin{tikzpicture}[baseline={([yshift=-.8ex]current bounding box.center)},scale=1.]
\draw [->-] (0, 0) -- (0, .7);
\draw [->-] (.5, 0) -- (.5, .7);
\draw [] (-.2, .7) -- (.7, .7) -- (.7, 1.3) -- (-.2, 1.3) -- (-.2, .7);
\draw [->-] (0, 1.3) -- (0, 2.);
\draw [->-] (.5, 1.3) -- (.5, 2.);
\node [centered, rotate=0.] at (-.3, .2) {\scriptsize $X_J$};
\node [centered, rotate=0.] at (.7, .2) {\scriptsize $y$};
\node [centered, rotate=0.] at (-.2, 1.8) {\scriptsize $y$};
\node [centered, rotate=0.] at (.8, 1.8) {\scriptsize $X_J$};
\node [centered, rotate=0.] at (.25, 1.) {\scriptsize $c_{X_J, y}$};
\end{tikzpicture}\qquad =\qquad
\begin{tikzpicture}[baseline={([yshift=-.8ex]current bounding box.center)},scale=1.]
\draw [->-] (1, 0) -- (1, 1);
\draw [->-] (1, 1) -- (1, 2);
\draw [->-] (.5, 0) arc [start angle = 180, end angle = 100, x radius = .5, y radius = 1.];
\draw [-<-] (1.5, 2) arc [start angle = 0, end angle = -80, x radius = .5, y radius = 1.];
\node [left, rotate=0.] at (0.5, .25) {\scriptsize $X_J$};
\node [right, rotate=0.] at (1.5, 1.75) {\scriptsize $X_J$};
\node [centered, rotate=0.] at (1.2, .2) {\scriptsize $y$};
\node [centered, rotate=0.] at (.8, 1.8) {\scriptsize $y$};
\end{tikzpicture}
}

\newcommand{\HalfBraidingB}{
\begin{tikzpicture}[baseline={([yshift=-.8ex]current bounding box.center)},scale=1]
\draw [->-] (1, 0) -- (1, 1);
\draw [->-] (1, 1) -- (1, 2);
\draw [->-] (.5, 0) arc [start angle = 180, end angle = 100, x radius = .5, y radius = 1.];
\draw [-<-] (1.5, 2) arc [start angle = 0, end angle = -80, x radius = .5, y radius = 1.];
\node [left, rotate=0.] at (0.5, .25) {\scriptsize $X_J$};
\node [right, rotate=0.] at (1.5, 1.75) {\scriptsize $X_J$};
\node [centered, rotate=0.] at (1.2, .2) {\scriptsize $y$};
\node [centered, rotate=0.] at (.8, 1.8) {\scriptsize $y$};
\end{tikzpicture}\qquad = \qquad \bigoplus_{p,q\in L_J}\ \bigoplus_{k\in L_\Fus}\ \sqrt{\frac{d_k}{d_y}}\ z_{pqy}^{J;k}\quad
\begin{tikzpicture}[baseline={([yshift=-.8ex]current bounding box.center)},scale=1]
\draw [->-] (1, 0) -- (1, .8);
\draw [->-] (1, 1.2) -- (1, 2);
\draw [->-] (1, .8) -- (1, 1.2);
\draw [->-] (.5, 0) arc [start angle = 180, end angle = 90, x radius = .5, y radius = .7];
\draw [-<-] (1.5, 2) arc [start angle = 0, end angle = -90, x radius = .5, y radius = .7];
\node [left, rotate=0.] at (0.5, .25) {\scriptsize $p$};
\node [right, rotate=0.] at (1.5, 1.75) {\scriptsize $q$};
\node [centered, rotate=0.] at (1.2, .2) {\scriptsize $y$};
\node [centered, rotate=0.] at (.8, 1.8) {\scriptsize $y$};
\node [centered, rotate=0.] at (1.2, 1.) {\scriptsize $k$};
\end{tikzpicture}
}

\newcommand{\HalfBraidingC}{
\begin{tikzpicture}[baseline={([yshift=-.8ex]current bounding box.center)},scale=1]
\draw [->-] (1.15, .4) arc (-60:30:.5);
\draw [->-] (.6, .1) -- (.9, .3);
\draw [->-] (1.2, 1.3) -- (1., 1.6);

\draw [->-] (1, 0) -- (1, 1);
\draw [->-] (.2, 1.5) -- (1, 1);
\draw [->-] (1.8, 1.5) -- (1, 1);
\node [centered, rotate=0.] at (1.15, 0.) {\scriptsize $y$};
\node [centered, rotate=0.] at (.2, 1.3) {\scriptsize $u$};
\node [centered, rotate=0.] at (1.8, 1.3) {\scriptsize $v$};
\node [centered, rotate=0.] at (.4, 0.) {\scriptsize $p$};
\node [centered, rotate=0.] at (.8, 1.7) {\scriptsize $q$};
\end{tikzpicture}\qquad = \qquad 
\begin{tikzpicture}[baseline={([yshift=-.8ex]current bounding box.center)},scale=1]
\draw [->] (.5, .5) arc (-170:-240:1.);
\draw [->] (.5, .5) arc (-170:-190:1.);
\fill [white] (.55, 1.3) circle (.2);

\draw [->-] (1, 0) -- (1, 1);
\draw [->-] (.2, 1.5) -- (1, 1);
\draw [->-] (1.8, 1.5) -- (1, 1);
\node [centered, rotate=0.] at (1.15, 0.) {\scriptsize $y$};
\node [centered, rotate=0.] at (.2, 1.3) {\scriptsize $u$};
\node [centered, rotate=0.] at (1.8, 1.3) {\scriptsize $v$};
\node [centered, rotate=0.] at (.5, .3) {\scriptsize $p$};
\node [centered, rotate=0.] at (1.1, 1.6) {\scriptsize $q$};
\end{tikzpicture}
}

\newcommand{\HalfBraidingD}{
\begin{tikzpicture}[baseline={([yshift=-.8ex]current bounding box.center)},scale=1]
\draw [->-] (1, 0) -- (1, 1);
\draw [->-] (1, 1) -- (1, 2);
\draw [->-] (.5, 0) arc [start angle = 180, end angle = 100, x radius = .5, y radius = 1.];
\draw [-<-] (1.5, 2) arc [start angle = 0, end angle = -80, x radius = .5, y radius = 1.];
\node [left, rotate=0.] at (0.5, .25) {\scriptsize $J$'s Charge $p$};
\node [right, rotate=0.] at (1.5, 1.75) {\scriptsize $J$'s Charge $q$};
\node [centered, rotate=0.] at (1.2, .2) {\scriptsize $j_E$};
\node [centered, rotate=0.] at (.8, 1.8) {\scriptsize $j_E$};
\end{tikzpicture} = \qquad
\sum_{k\in L_\Fus} \sqrt{\frac{d_k}{d_{j_E}}}\ \overline{z_{pqj_E}^{J;k}} 
\begin{tikzpicture}[baseline={([yshift=-.8ex]current bounding box.center)},scale=1]
\draw [->-] (1, 0) -- (1, .8);
\draw [->-] (1, 1.2) -- (1, 2);
\draw [->-] (1, .8) -- (1, 1.2);
\draw [->-] (.5, 0) arc [start angle = 180, end angle = 90, x radius = .5, y radius = .7];
\draw [-<-] (1.5, 2) arc [start angle = 0, end angle = -90, x radius = .5, y radius = .7];
\node [left, rotate=0.] at (0.5, .25) {\scriptsize $p$};
\node [right, rotate=0.] at (1.5, 1.75) {\scriptsize $q$};
\node [centered, rotate=0.] at (1.2, .2) {\scriptsize $j_E$};
\node [centered, rotate=0.] at (.8, 1.8) {\scriptsize $j_E$};
\node [centered, rotate=0.] at (1.2, 1.) {\scriptsize $k$};
\end{tikzpicture}\qquad =\qquad
W_E^{J;pq}\quad
\begin{tikzpicture}[baseline={([yshift=-.8ex]current bounding box.center)},scale=1]
\draw [->-] (1, 0) -- (1, 2);
\node [centered, rotate=0.] at (1.3, 1.) {\scriptsize $j_E$};
\end{tikzpicture}\ .
}

\begin{document}

\title{Nonabelian Anyon Condensation in 2+1d topological orders: A String-Net Model Realization}

\date{\today}
\author[a,b]{Yu Zhao}
\author[a,b]{Yidun Wan\footnote{Corresponding author}}
\affiliation[a]{State Key Laboratory of Surface Physics, Center for Astronomy and Astrophysics, Department of Physics, Center for Field Theory and Particle Physics, and Institute for Nanoelectronic devices and Quantum Computing, Fudan University, 2005 Songhu Road, Shanghai 200433, China}
\affiliation[b]{Shanghai Research Center for Quantum Sciences, 99 Xiupu Road, Shanghai 201315, China}
\emailAdd{yuzhao20@fudan.edu.cn, ydwan@fudan.edu.cn}

\abstract{We develop a comprehensive framework for realizing anyon condensation of topological orders within the string-net model by constructing a Hamiltonian that bridges the parent string-net model before and the child string-net model after anyon condensation. Our approach classifies all possible types of bosonic anyon condensation in any parent string-net model and identifies the basic degrees of freedom in the corresponding child models. Compared with the traditional UMTC perspective of topological orders, our method offers a finer categorical description of anyon condensation at the microscopic level. We also explicitly represent relevant UMTC categorical entities characterizing anyon condensation through our model-based physical quantities, providing practical algorithms for calculating these categorical data.}
\maketitle

\flushbottom

\section{Introduction}\label{sec:intro}

A complete understanding of topological orders urges the study of topological phase transitions. Anyon condensation\cite{Bais2009a, Barkeshli2010, Burnell2012,schulz2013, Barkeshli2013a, Eliens2013, Chen2013, Gu2014a, HungWan2015a, Ji2019, Hu2021, xu2021, xu2022} is a main mechanism triggering topological phase transitions. In this article, we focus on bosonic anyon condensation, where all condensed anyons are self-bosons with trivial self-statistics and mutual statistics. Abstractly in the categorical description, the topological properties of a topological order are characterized by a unitary modular tensor category (UMTC), while a set of condensable anyons is described by a \emph{commutative separable Frobenius algebra} (CSFA) object in the UMTC\cite{HungWan2015a, Wan2017}: 
\eq{\text{Parent Order}\quad
\text{\raisebox{5pt}{$\underrightarrow{\text{\scriptsize\ \ \ \ CSFA \name{A}\ \ \ \ }}$}}\quad\text{Child Order}.}
When the anyons in the CSFA condense, the corresponding (parent) topological order will undergo a phase transition and become another (child) topological order, whose topological properties are characterized by a smaller UMTC. 

While the UMTC categorical perspective provides critical insights into anyon condensation of topological orders, it primarily focuses on the mathematical relations between the topological properties of the parent and child order and overlooks specific details during phase transition, particularly (1) the dynamics of the phase transition, (2) how anyons are modified in a physical context, and (3) the transformations of the basic degrees of freedom (dofs) that constitute the field configuration. These microscopic details are central to traditional phase transition theories like the Higgs mechanism and are crucial in dynamically analyzing phase transitions, including understanding the critical behavior and determining the universality class. Since critical points lie in a non-topological region, intermediate processes during phase transitions cannot be captured in the categorical description.

To fill this gap, exactly solvable models, such as the string-net model\cite{Bravyi1998, Kitaev2003a, Levin2004, Buerschaper2009, Hung2012, Hu2012, Hu2012a, Buerschaper2013, schulz2013, Lan2014b, Hu2017, Hu2018, cheipesh2019, Wang2020, Wang2022, zhao2022, zhao2024}, can be used to represent how basic dofs of topological orders transform because anyon excitations are represented as concrete excited states composed of basic dofs in the model's Hilbert space. The basic dofs of the parent string-net model form a unitary fusion category (UFC), while those of the child model form a subcategory of the parent UFC. Previous works\cite{zhao2022, lin2023} have realized Abelian anyon condensation in the string-net model. Nevertheless, general non-Abelian anyon condensation remains an open question in the string-net model.

In this paper, we solve this question by systematically constructing generic anyon condensation in the string-net model and obtain the following key results:

\begin{enumerate}
\item A practical recipe for realizing all possible bosonic anyon condensations within topological orders describable by string-net model, and identify the basic dofs of the resultant child string-net model. A novel finer categorical framework of anyon condensation, compared with the UMTC description, is extracted from our recipe.
\item An exact connection between our model-based anyon condensation recipe and the UMTC description of anyon condensation---a CSFA in the UTMC of the parent order. We prove that the CSFA is the full center of the input UFC of the child model within the input UFC of the parent model. We provide an algorithm based on our recipe to calculate this CSFA.
\end{enumerate}

Our approach is fully physics-based and computational, although categorical language may be used at certain points only for corroboration and comparison. To elaborate, bosonic anyon condensation implies that the new vacua of the child order are coherent states containing arbitrarily many condensed bosons from the parent order\cite{Hu2021, zhao2022}, analogous to the Cooper pair condensation in superconductivity phase transitions. To yield such coherent states, we add to the Hamiltonian of the parent model a sum of creation operators for the condensed anyons, with specific coefficients to make the addent a projector \(P\):
\eqn[eq:intro]{
H_{\rm Parent} \qquad\Longrightarrow\qquad H_{\rm Parent} - \lim_{\Lambda\to\infty}\Lambda P .}
In the limit \(\Lambda \to \infty\), the projector \(P\) (defined in Eq. \eqref{eq:CondenseHamil}) ensures that the new ground states are \(+1\) eigenstates of \(P\), which are coherent states filled with arbitrarily many condensed anyons throughout the lattice.
\begin{figure}[!h]
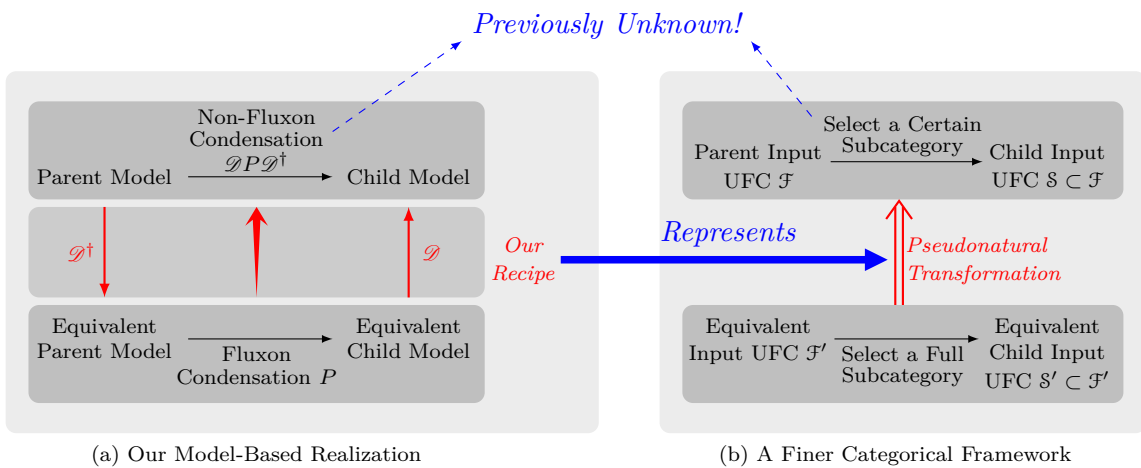
\centering
{\Sketch}
\caption{Extracted from (a) Our recipe of non-fluxon condensation in the string-net model constructed by duality map \(\D\) is (b) A finer categorical framework of anyon condensation. Mathematically, (a) is a representation of (b).}
\label{fig:intro}
\end{figure}

The remaining challenge is to explicitly construct the projector \(P\), such that it describes an anyon condensation process. In the simplest cases discussed in Section \ref{sec:anyoncond}, projector \(P\) projects out several basic dofs of the parent string-net model, such that the remaining dofs serve as the basic dofs of the child model. The input UFC of this child model is a \emph{full subcategory} of the parent input UFC. Such anyon condensation is referred to as \emph{fluxon condensation}, which was preliminarily studied in our previous work\cite{zhao2022}. Fluxon condensation must be bosonic because all fluxons are bosons. Nevertheless, anyon condensation is in general not fluxon condensation. Section \ref{sec:generalcond} details our recipe for constructing non-fluxon condensation by leveraging fluxon condensation. Our previous work\cite{zhao2024} classified all equivalent string-net models describing the same topological order and established unitary duality maps among these models. These duality maps may transform fluxons into non-fluxons. Therefore, we can obtain a non-fluxon bosonic condensation projector \(P\) \eqref{eq:intro} for any given parent model by transforming a fluxon condensation projector in an equivalent model back to the parent model by the duality map \(\D\) between the two models:
\eqn[eq:introdual]{P = \D P_{\rm Dual}\D^\dagger,}
where \(\D\) is the duality map, and \(P_{\rm Dual}\) is a fluxon condensation projector in the dual model that is easier to calculate. Figure \ref{fig:intro}(a) depicts our recipe as commutative diagrams. By applying all possible duality maps for all types of fluxon condensations in all equivalent models, we obtain all possible bosonic anyon condensation in a given string-net model.

In contrast to the UMTC description of anyon condensation at the output level of the string-net model, a finer categorical framework for anyon condensation at the input UFC level is extracted from our recipe, as shown in Fig. \ref{fig:intro}(b). Let's be concrete. A parent string-net model is specified by an input UFC $\Fus$, while one of its equivalent model is specified by input UFC $\Fus'$. A fluxon condensation in such an equivalent model is mathematically a functor that selects in $\Fus'$ a full subcategory $\Child'$ as the input UFC of the child string-net model after the fluxon condensation. The duality operator $\D$ that sends this fluxon condensation to a certain non-fluxon condensation in the original parent string-net model is mathematically a \textit{pseudonatural transformation}\cite{kachour2016, corner2017, cheipesh2019}: $\D$ maps parent $\Fus'$ to $\Fus$ and child $\Child'$ to $\Child$ at the input UFC level, while the functor selecting $\Child'\subset \Fus'$ to that selecting $\Child\subset \Fus$. This new, finer categorical framework involving pseudonatural transformations is abstract, but our recipe involving duality maps is computable in the string-net model's Hilbert space, as duality maps are concrete matrices in terms of the model's basic dofs. Therefore, our recipe literally represents this finer categorical framework. In particular, our duality operator $\D$ represents a pseudonatural transformation. 

Note that this finer categorical framework of anyon condensation is completely defined at a microscopic level in terms of the basic dofs of the string-net model and cannot be observed in the UMTC description. A duality map \(\D\) preserves the output UMTC and hence, the CSFA \(\name{A}\) characterizing an anyon condensation is invariant under the duality maps, which only changes how \(\name{A}\) is represented by different input UFCs. Hence, the traditional UMTC categorical description overlooks the finer structures of anyon condensation.

We also discover that the CSFA \(\name{A}\) associated with a bosonic anyon condensation is the \emph{full center}\cite{fjelstad2008, kong2008, Kong2013} of the child UFC \(\Child\) within the parent UFC \(\Fus\). Explicitly calculating the full center for a given subcategory in a UFC is generally challenging in category theory. Our recipe of anyon condensation in the string-net model offers a physical approach to tackle this mathematical challenge. In Section \ref{sec:category}, we demonstrate that the operator expansion coefficients in the summands of a projector \(P\) \eqref{eq:intro} directly represent the CSFA \(\name{A}\) of the corresponding anyon condensation.

Our systematic construction of anyon condensation paves the way for a deeper exploration of anyon-condensation-induced phase transitions and the corresponding symmetry breaking---a generalized version beyond the conventional Landau-Ginzburg paradigm\cite{Hung2013, Hu2021}. In previous works, numerical methods like tensor networks \cite{xu2021, xu2022} have uncovered several critical behaviors in certain topological phase transitions. Nevertheless, in these tensor-network approaches, the evolution is governed by transfer matrices, hence, the microscopic dynamics---including how the fundamental degrees of freedom, anyon types, and the Hamiltonian's energy spectra evolve during anyon condensation---are not captured. In contrast, our recipe directly controls the Hamiltonian, formulating a physical evolution of the system's phase: By gradually increasing the parameter \(\Lambda\) in Eq. \eqref{eq:intro} from \(0\) to \(\infty\), we can model a physical continuous phase transition process from the parent order to the child order. Besides, in our framework, different types of anyon condensation can lead to distinct child string-net models with varying basic dofs, all of which describe the same child order. These different anyon condensation processes may not be distinguishable within the categorical description (as seen in the example of the doubled Fibonacci topological order in the main text), only becoming discernible when analyzed through a model. Yet, such differences are reasonable because a child order always possesses a global symmetry arising from breaking the symmetry of the parent order. Different child string-net models of the same child order are different symmetry sectors of this global symmetry. This insight suggests a pathway for modeling the symmetry transformations of symmetry-protected topological (SPT) orders\cite{Chen2011e, Chen2011d, Oon2012, Hung2012b, Chen2013, Lichtman2020, seifnashri2024} and symmetry-enriched topological (SET) orders\cite{Mesaros2011, Hung2013, Gu2014a, Hu2021}. These works are ongoing and are to be reported elsewhere.

\section{Brief Review of String-Net Model and Anyon Condensation}\label{sec:model}

In this section, we briefly review the string-net model and phenomena in anyon condensation. 
A more detailed review of the string-net model is presented in Appendix \ref{sec:review}.

\subsection{String-Net Model}

\begin{figure}[!ht]
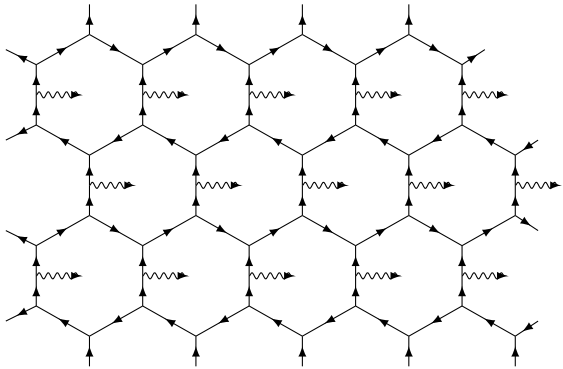
\centering
\Lattice
\caption{Part of the string-net model lattice. A tail (wavy line) is attached to an arbitrary edge of every plaquette.}
\label{fig:lattice}
\end{figure}

We take the form of the string-net model defined in Ref. \cite{zhao2024}, which was adapted from that in \cite{Hu2018}. The string-net model is an exactly solvable Hamiltonian model defined on a \(2\)-dimensional oriented trivalent lattice, e.g., that depicted in Fig. \ref{fig:lattice}. Each plaquette has a tail attached to one of its edges\footnote{The original string-net model in Ref. \cite{Levin2004}, which lacks these tails, cannot fully describe charge excitations. These added tails carry the charges of anyons, thereby enlarging the Hilbert space to encompass the complete anyon spectrum.}. Different choices of the edge to which the tail is attached are equivalent, meaning that the tail can be moved freely from one edge to another while keeping the physical states invariant, as demonstrated in Appendix \ref{sec:pachner}.

The input data of the string-net model is a \emph{unitary fusion cateogry} (UFC) \(\Fus\), described by a finite set \(L_\Fus\) equipped with three functions: the \emph{fusion rules} \(\delta: L_\Fus^3 \to \NN\), the \emph{quantum dimension} \(d: L_\Fus \to \RR\), and the $6j$-\emph{symbol} \(G: L^6_\Fus \to \C\) (see Appendix \ref{sec:review}). The elements in \(L_\Fus\) are the representative \emph{simple objects} of \(\Fus\). The basic configuration of the string-net model is established by labeling each edge and tail with an element in \(L_\Fus\), subject to the constraint that \(\delta_{abc} \ne 0\) for any three incident edges or tails counterclockwise carrying dofs \(a, b, c \in L_\Fus\), meeting at a vertex and all pointing toward this vertex. We can reverse the direction of any edge or tail and simultaneously conjugate its label as \(a \to a^\ast\), the opposite simple object of \(a\in L_\Fus\), which keeps the configuration invariant. The Hilbert space \(\Hil_\Fus\) is spanned by all possible such configurations.

The Hamiltonian reads
\eqn[eq:Hamil]{H_\Fus := - \sum_{{\rm Plaquettes}\ P}Q_P^\Fus.}
The plaquette operators \(Q_P^\Fus\) are commuting projectors detailed in Appendix \ref{sec:review}, making the Hamiltonian \(H_\Fus\) exactly solvable. The ground states \(\ket\Phi_\Fus\) are common $+1$ eigenstates of all \(Q_P^\Fus\) operators. An excited state \(\ket\psi_\Fus\) is another common eigenstate that satisfies \(Q_P^\Fus\ket{\psi}_\Fus = 0\) for one or more plaquettes \(P\), each of which bears an \emph{anyon}. We also say a ground
state has a trivial anyon \(\trivial\) in each plaquette. Anyon species \(J\) takes value as a simple object in \(\Fus\)'s \emph{Drinfeld center} \(\Cent(\Fus)\), a UMTC encapsulating the topological properties of the output topological order described by the string-net model. 

The advantage of describing a topological order using the string-net model lies in its ability to manifest the \emph{internal spaces} of anyons\cite{zhao2022, zhao2023, zhao2024}. In the string-net model, an excited state is determined not only by the anyon species in each plaquette but also by the anyon's internal charge---the dof on the tail where the anyon resides. In other words, the string-net model represents an anyon \(J\) as a \emph{dyon}, a pair \((J, p)\) comprising the anyon's species \(J\) and its internal charge \(p\in L_\Fus\) on the tail. A non-Abelian anyon may carry multiple charge types and is thus represented within a multi-dimensional Hilbert subspace of excited states in the model, in contrast to the categorical description of topological order, where an anyon is labeled by a simple object that is an indecomposable element of \(\Cent(\Fus)\). The internal space of an anyon, expanded by its charges, serves as the representation space of the gauge symmetry of the topological order\cite{zhao2024}. 

Another advantage is that the string-net model allows to study the phase transitions of topological orders in line with the conventional field-theoretical methods because the string-net model can be either understood as an effective lattice gauge theory\cite{zhao2024} or a microscopic model of topological orders. By evoluting the Hamiltonian \eqref{eq:Hamil}, one may identify the critical points, analyze critical behaviors, and determine universality classes of phase transitions. The categorical perspective fails to capture these critical properties because the critical points lie in a non-topological region.

Let's consider a simple but archetypal example: the Fibonacci string-net model describing the doubled Fibonacci topological order. The input Fibonacci UFC \(\Fibo\) has two simple objects \(1\) and \(\tau\). The Hilbert space \(\Hil_\Fibo\) is spanned by all possible assignments of \(1\) and \(\tau\) to all edges/tails, dictated by the nonzero fusion rule \(\delta_{abc} = 1\) for three edges/tails \(a, b, c\) meeting at any vertex, where
\eq{
\delta_{111} = \delta_{1\tau\tau} = \delta_{\tau 1\tau} = \delta_{\tau\tau 1} = \delta_{\tau\tau\tau} = 1,\qquad \delta_{11\tau} = \delta_{1\tau 1} = \delta_{\tau 11} = 0.
}
The output doubled Fibonacci topological order has four anyon species, labeled by simple objects of UMTC \(\Cent(\Fibo)\):
\eq{1\bar 1,\qquad \tau\bar 1,\qquad 1\bar\tau,\qquad\tau\bar\tau,
}
where \(1\bar 1\) labels the trivial anyon. Fibonacci string-net model has five dyon types \((J, p)\): 
\eq{
(1\bar 1, 1),\qquad (1\bar\tau, \tau),\qquad (\tau\bar 1, \tau),\qquad (\tau\bar\tau, 1),\qquad (\tau\bar\tau, \tau).}
An anyon \(\tau\bar\tau\) can have a charge of either \(1\) or \(\tau\) in the model.

\subsection{Anyon Condensation}\label{sec:condreview}

We now briefly review the physical phenomena that occur during phase transitions triggered by a bosonic anyon condensation, as well as their categorical descriptions.

Physically, after anyon condensation, the new vacuum states of the child order become coherent states containing arbitrarily many condensed anyons from the parent order, analogous to the Cooper-pair condensation in superconductivity phase transitions and Higgs boson condensation in the standard model. Several phenomena arise due to this condensation\cite{Bais2009, Gu2014a, Hu2021, zhao2022}:
\begin{enumerate}
\item \emph{Splitting}: During anyon condensation, certain anyons, including those condensed, may split into multiple sectors that become distinct anyon species in the child order. That is, a condensed anyon might not fully condense but may only partially condense into one or more sectors, akin to the Higgs boson condensation in electroweak symmetry breaking, where only a one-dimensional subspace of the two-dimensional Higgs boson space condenses. 
    
\item \emph{Identification}: As condensed sectors become the new vacuum, two types of topological sectors related by fusing with a condensed sector in the parent order can no longer be distinguished in the child order and are therefore identified as the same type of child anyon.

\item \emph{Confinement}: Not all anyons are free in the child order. Anyons braiding nontrivially with the condensate become confined in the child order because the new vacuum should not be disturbed by moving anyons around. In other words, creating confined anyons requires overcoming an infinitely large energy barrier. This is analogous to magnetic-flux confinement in Copper-pair condensation.
\end{enumerate}

In the categorical description of topological order theory, where the topological properties of the parent order are encapsulated by a UMTC \(\mathcal{C}\) and anyons are labeled by simple objects in \(\mathcal{C}\), a commutative separable Frobenius algebra (CSFA) \(\name{A}\)\cite{HungWan2015a, Wan2017}---a composite object in \(\mathcal{C}\)---is introduced to describe anyon condensation. The term ``commutative'' refers to the requirement that all condensed anyons are bosons with trivial self-statistics and mutual braidings with each other, as they become the vacuum in the child order. The term ``Frobenius algebra'' indicates that fusing two CSFA objects \(\name{A}\) is isomorphic to \(\name{A}\) itself, representing the trivial fusion rule of the vacuum in the child order. The condition ``separable'' ensures that the representations over \(\name{A}\) in \(\mathcal{C}\) form a fusion category \(\rep_{\mathcal{C}}\name{A}\), whose simple objects are irreducible representations over \(\name{A}\), and CSFA \(\name{A}\) itself is the trivial representation over \(\name{A}\).

\begin{figure}[!ht]
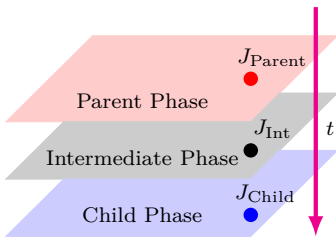
\centering
\FigureCorrespondence
\caption{An auxiliary intermediate order (gray sheaf) is introduced during anyon condensation between the parent order (red sheaf) and child order (blue sheaf) to analyze the process of anyon condensation.}
\label{fig:intermediate}
\end{figure}

An auxiliary intermediate order is introduced as a method to anatomize the procedure of anyon condensation\cite{Bais2009a, zhao2022}, see Fig. \ref{fig:intermediate}. The anyon condensation first leads to the intermediate order where splitting and identification have been completed. This intermediate order is encapsulated by \(\rep_{\mathcal{C}}\name{A}\)\cite{HungWan2015a, Wan2017}. Quasiparticle sectors in this intermediate order are labeled by simple objects in \(\rep_{\mathcal{C}}\name{A}\). Each irreducible representation is a composite object in \(\mathcal{C}\). Two anyons species \(J, K \in L_{\mathcal{C}}\) appearing in the same object in \(\rep_{\mathcal{C}}\name{A}\) are identified, while a splitted anyon species \(J\) appears in more than one object in \(\rep_{\mathcal{C}}\name{A}\).

This intermediate order is generally not the child order because \(\rep_{\mathcal{C}}\name{A}\) lacks a well-defined braiding structure. So it is a pure auxiliary mathematical tool without physical meaning. Some quasiparticle sectors in the intermediate order braid nontrivially with the new vacuum, leading to their confinement in the child order. The new vacuum is labeled by the trivial representation \(\name{A}\). Therefore, \(\rep_{\mathcal{C},0}\name{A}\)---a subcategory of \(\rep_{\mathcal{C}}\name{A}\) consisting of those representation objects that have trivial braiding with \(\name{A}\)---is a UMTC with a well-defined braiding structure and characterizes the topological properties of the child order.

One should note that different sectors of an anyon after splitting are physically meaningful, analogous to the Higgs boson condensation triggering electroweak symmetry breaking, where only one real component of the Higgs complex-doublet condenses. Nevertheless, they are not captured by the UMTC categorical description. Anyons are represented as simple objects in the parent UMTC, which does not account for finer internal structures in the categorical language. To address these splitted sectors, one must consider the internal spaces of an anyon---the internal charges in a given string-net model. In this context, it becomes meaningful to discuss the behaviors of anyons during splitting phenomena.

Consider anyon condensation in the doubled Fibonacci topological order as an example. In the categorical description of this topological order, there is a unique anyon condensation that partially condenses anyon \(\tau\bar\tau\), which completely breaks the doubled Fibonacci topological order to the trivial topological order:
\begin{enumerate}
\item Splitting: \(\tau\bar\tau\) formally splits to two sectors \(1\) and \(\chi\). These two sectors have no meaning within the doubled Fibonacci UMTC. Instead, we will later establish their significance within the string-net model.
\item Identification: In the intermediate order, \(1\bar 1\) is identified with \(1\); \(\tau\bar 1\) and \(1\bar\tau\) are identified with \(\chi\).
\item Confinement: \(\chi\) is confined in the child phase.
\end{enumerate}
Now that anyon species \(1\bar 1\) and \(\tau\bar\tau\) are condensed, the CSFA \(\name{A}\) describing \(\tau\bar\tau\) anyon condensation is a composite object \(1\bar 1\oplus\tau\bar\tau\) in UMTC \(\Cent(\Fibo)\). In Section \ref{sec:category}, we'll use our method to derive the explicit algebraic structure of \(\name{A}\).

Hereafter, we focus on doubled topological orders, which are those describable by the string-net model.

\section{Generic Anyon Condensation in the String-net Model}\label{sec:anyoncond}

We now discuss how to establish anyon condensation in a parent string-net model with an input UFC \(\Fus\), which describes the parent order before anyon condensation. 

To realize the coherent states after anyon condensation, we propose adding an \emph{anyon condensation term} to the parent Hamiltonian \(H_\Fus\):
\eqn[eq:EffHamil]{
H_\Fus \to H_\Fus -\lim_{\Lambda\to\infty}\Lambda\sum_{\text{Edges } E} P_E,}
\eqn[eq:CondenseHamil]{
P_E := \sum_{\text{Condensed Anyons } J}\sum_{J\text{'s Charges }p,q}\frac{\pi_J^{pq}}{d_pd_q} W_E^{J; pq}.
}
Here, \(d_p\) (\(d_q\)) is the quantum dimension of charge \(p\) (\(q\)), and \(W_E^{J;pq}\) is the simplest creation operator that creates a pair of dyons \((J^\ast, p^\ast)\) and \((J, q)\) in two \emph{adjacent} plaquettes across an edge \(E\):
\eq{
W_E^{J; pq}\quad \ExcitedSrc\quad := \quad \sum_{k \in L_\Fus} \cdots\ \ExcitedTar\ ,
}
where \(j_E\in L_\Fus\) is the dof on edge \(E\). We only present the specific case where the two tails in the adjacent plaquettes are both \(1\) and omit the expansion coefficients denoted by ``\(\cdots\)'' here for brevity and clarity. The explicit matrix elements are detailed in Appendix \ref{sec:spec}. These simplest creation operators are sufficient to define the condensation term because all other creation operators are compositions of them (see Appendix \ref{sec:spec} for details). The condensed dyons refer to those dyons \((J, p)\) satisfying \(\pi_J^{pq}\ne 0\) for some charge \(q\). The coefficients \(\pi_J^{pq}\) are chosen to make \(P_E\) \eqref{eq:CondenseHamil} a \emph{projector}. For \(\Lambda \to \infty\), projector \(P_E\) ensures that the new ground states are \(+1\) eigenstates of \(P_E\)---the sum of condensed anyon’s creation operators \(W_E^{J;pq}\), making the new ground states coherent states with arbitrarily many condensed anyons \(J\) throughout the lattice. The resultant child order after anyon condensation is described by the child string-net model, with the child Hilbert space \(\Hil_{\rm Child}\) and child Hamiltonian \(H_{\rm Child}\) obtained by applying projector \(P_E\) to those in the parent model:
\eqn[eq:ChildModel]{
\Hil_{\rm Child} = \Bigg[\prod_{\text{Edges }E} P_E\Bigg]\Hil_\Fus,\qquad H_{\rm Child} = \Bigg[\prod_{\text{Edges }E} P_E\Bigg]H_\Fus\Bigg[\prod_{\text{Edges }E} P_E\Bigg].
}
Hamiltonian \(H_{\rm Child}\) is exactly solvable up to an irrelevant global scalar factor arising from the projection. The spectrum of the child model is directly determined by projector \(P_E\) \eqref{eq:CondenseHamil}, which projects the ground states \(\ket\Phi_\Fus\), excited states \(\ket\varphi_\Fus\), and creation operators \(W_E^{J;pq}\) of the parent model into those of the child model\cite{zhao2022}. 

The phenomena associated with anyon condensation can be characterized  in the string-net model as follows:
\begin{enumerate}
\item \emph{Splitting} happens when two dyons \((J, p)\) and \((J, q)\) with \(p\ne q\) are projected to distinct anyon species in the child model. In this case, the internal space of the parent anyon ``splits'' into the direct sum of the internal spaces of distinct child anyons.
\item \emph{Identification} occurs when two parent anyon species are projected onto the same anyon species in the child model.
\item \emph{Confinement}: Excited states of the parent model involving anyons to be confined in the child order are projected out of the child model's Hilbert space \(H_{\rm Child}\).
\end{enumerate}

Two problems remain: How to calculate the coefficients \(\pi_J^{pq}\), such that \(P_E\) is a projector representing an anyon condensation process, and what is the child input UFC of the child model? 

\section{Ease with Fluxon Condensation}

The remaining problems mentioned above can be easily addressed in the case of \emph{fluxon condensation}. A fluxon refers to a dyon with a trivial charge \(1\in L_\Fus\) in the string-net model, inducing that each basis state of the string-net model is an eigenstate of the fluxon creation operator: 
\eqn[eq:FluxCreate]{W_E^{J;11}\quad \ExcitedFlux\quad =\quad w_J(j_E)\quad\ExcitedFlux\ ,}
where \(j_E\) is the dof on edge \(E\), and \(w_J(j_E)\in\CC\). To do fluxon condensation, we first choose a subset \(L_\Sub \subset L_\Fus\) of simple objects that are closed under fusion. The subset \(L_\Sub\) generates a \emph{full subcategory} \(\Sub\) of the parent input UFC \(\Fus\). This subcategory \(\Sub\) is equipped with the same \(\delta\), \(d\), and \(G\) functions as those of \(L_\Fus\) but restricted to \(L_\Sub\). As \(\Sub\) is also a UFC, it can serve as the input UFC for a new string-net model, referred to as the \emph{child string-net model}, whose basic dofs on edges and tails take values from the simple objects in subset \(L_\Sub\). The transition from the parent model to the child model involves gapping out those dofs that are not in \(L_\Sub\), such that the condensation projector \(P_E\) \eqref{eq:CondenseHamil} becomes
\eqn[eq:FluxonCondensation]{
P_E^{\Sub|\Fus}\ket{\psi} = \Bigg[\sum_{\text{Condensed Anyons }J} \pi_J^{11} w_J(j_E)\Bigg]\ket{\psi} = \delta_{j_E \in L_\Sub}\ket{\psi},}
where \(\delta\) is the Kronecker symbol, and \(j_E\) is the dof on edge \(E\) in basis state \(\ket\psi\). Operator \(P_E^{\Sub|\Fus}\) is manifestly diagonal due to Eq. \eqref{eq:FluxonCondensation}. The number of fluxon species in a string-net model always equals the number of simple objects\cite{Hu2018}, so Eq. \eqref{eq:FluxonCondensation} always has a unique solution. The anyon species \(J\) satisfying \(\pi_J^{11} \ne 0\) are condensed during fluxon condensation, resulting in a child order described by a child model with the child input UFC \(\Sub\). The parent dyons \((J, p)\) with \(p\notin L_\Sub\) are confined in the child model. The topological properties of the child order are encapsulated by UMTC \(\Cent(\Sub)\).

All types of fluxon condensation are bosonic because fluxons must be bosons with trivial self- and mutual statistics. This is reflected in the fact that the creation operators \(W_E^{J;11}\) for fluxons are commutative. This property is apparent, as fluxon creation operators \(W_E^{J;11}\) are diagonal matrices.

In the Fibonacci string-net model, which has two fluxon species \((1\bar 1, 1)\) and \((\tau\bar\tau, 1)\), the creation operator \(W_E^{\tau\bar\tau; 11}\) has eigenvalues
\[
w_{\tau\bar\tau}(1) = 1,\qquad w_{\tau\bar\tau} (\tau) = - \frac{1}{\phi^2},\qquad \phi = \frac{\sqrt{5} + 1}{2},
\]
which induces a unique fluxon condensation projector
\eq{
P_E^{\Sub_0|\Fibo} = \pi_{1\bar 1}^{11}W_E^{1\bar 1; 11} + \pi_{\tau\bar\tau}^{11}W_E^{\tau\bar\tau;11} =
\frac{\idm + \phi^2W_E^{\tau\bar\tau; 11}}{\phi^2 + 1},}
where 
\eq{\pi_{1\bar 1}^{11} = \frac{1}{\phi^2 + 1},\qquad \pi_{\tau\bar\tau}^{11} = \frac{\phi^2}{\phi^2 + 1}.}
The fluxon condensation projector gaps out the nontrivial dof \(\tau\) on edge \(E\) in basis \(\ket\psi\):
\eq{P_E^{\Sub_0|\Fibo}\ket\psi = \delta_{j_E,1}\ket\psi.}
The child input UFC of the child order is the trivial UFC \(\Sub_0\) with \(L_{\Sub_0} = \{1\}\). The child Hilbert space is one-dimensional because each edge or tail is assigned with \(1\). Therefore, the resultant child model describes a trivial topological order, consistent with the known fact that condensing anyon \(\tau\bar\tau\) in the doubled Fibonacci topological order breaks the topological order completely. It is argued in Ref. \cite{Hung2012, Hung2013, HungWan2013a, Gu2014a} that this trivial topological order possesses a global symmetry that is algebraic or categorical, so it is an algebraic symmetry-protected topological phase. We will explore this intriguing scenario in an ongoing work.

\section{Difficulties with Dyon Condensation and Our Solution}\label{sec:generalcond}

While the two problems brought up at the end of Section \ref{sec:anyoncond} can be easily addressed in the case of fluxon condensation, they become difficult in the case of non-fluxon condensation, viz generic dyon condensation.  In other words, the coefficients \(\pi_{J}^{pq}, p,q \in L_\Fus\) in Eq. \eqref{eq:CondenseHamil} are generally difficult to calculate in the case of dyon condensation. Furthermore, the child input UFC \(\Child\) of the resultant child string-net model after dyon condensation becomes obscure. Since the projector \(P_E^{\Child|\Fus}\) no longer simply gaps basic dofs in the parent model, the child dofs are not apparent and do not exhibit a straightforward UFC structure.

We now tackle these difficulties by applying \emph{duality maps} between different but equivalent string-net models describing the same topological order, as developed in our previous work\cite{zhao2024}. Our duality maps dofs in one model to superpositions of those in another model, thereby transforming excited states with fluxons only in one model into superpositions of excited states with dyons in an equivalent model. Building on this, we offer a recipe to realize dyon condensation in a given string-net model with input UFC \(\Fus\):
\begin{enumerate}
\item Select a dual model with input UFC \(\Fus'\) of the original string-net model. Let \(\D\) be the duality map from the dual model to the original model.
\item Perform a fluxon condensation in this dual model by constructing the fluxon condensation term \(P_E^{\Sub'|\Fus'}\), where \(L_{\Sub'} \subset L_{\Fus'}\).
\item Transform the fluxon condensation term \(P_E^{\Sub'|\Fus'}\) from the dual model back into the original model by the duality map \(\D\), resulting in a dyon condensation term in the original model:
\eqn[eq:DyonCondensation]{
P_E^{\Child|\Fus} = \D P_E^{\Sub'|\Fus'} \D^\dagger,}
where \(\Child\) is the child input UFC of the child string-net model after dyon condensation in the original parent model.
\end{enumerate}

Equation \eqref{eq:DyonCondensation} directly calculate the coefficients \(\pi_J^{pq}\) of dyon condensation projector \eqref{eq:CondenseHamil} as combinations of the coefficients of fluxon condensation term \(P_E^{\Sub'|\Fus'}\) in the dual model, bypassing the difficulty of solving the projector condition for these coefficients. The full subcategory \(\Sub'\) generated by \(L_{\Sub'} \subset L_{\Fus'}\) in the dual model is also transformed into an isomorphic, yet less apparent, subcategory \(\Child\) of \(\Fus\), solving the difficulty of determining the basic dofs and input UFC of the child model. Given an original string-net model, by applying duality maps to different fluxon condensations in different dual models, we can generate \emph{all} bosonic dyon condensation within the original string-net model. We expound on our recipe in what follows.

\subsection{Duality Map}\label{sec:dualitymap}

A UFC \(\Fus\) includes \emph{Frobenius algebra} objects. According to a mathematical theorem\cite{etingof2016}, the bimodules---a special class of representations---of a given Frobenius algebra \(\A\) in \(\Fus\) form another UFC \(\Bimod_\Fus(\A)\) that is \emph{categorically Morita equivalent} to \(\Fus\). The dual string-net model with Morita equivalent input UFC \(\Bimod_\Fus(\A)\) is equivalent to the original string-net model with input UFC \(\Fus\), in the sense that they describe the same topological order whose topological properties are encapsulated by
\eqn[eq:Equivalence]{
\Cent(\Bimod_\Fus(\A)) = \Cent(\Fus).
}
Conversely, any UFC categorically Morita equivalent to \(\Fus\) can be expressed as a bimodule category over certain Frobenius algebra in \(\Fus\). Based on these mathematical facts, in Ref. \cite{zhao2024}, we classified all equivalent string-net models describing the same topological order and established duality maps \(\D\) that transform the basic dofs on edges and tails of one model to superpositions of the basic dofs of another model.

The algebra space of a Frobenius algebra \(\A\) is spanned by basis elements labeled by \(a_i\), where \(a \in L_\Fus\), \(i = 1, 2, \cdots, n_a^\A\) are \emph{multiplicity labels}, and \(n_a^\A\) is the multiplicity of \(a\) appearing in \(\A\). Different \(a_i\) are all simple object \(a\) but regarded as distinct basis elements because they follow different algebra multiplication rules, encoded in a function \(f: L_\A^3 \to \mathbb{C}\), where \(L_\A\) is the set of all basis elements \(a_i\). Specifically, Frobenius algebra \(\A\) is given by
\[
\A = \left\{\sum_{a \in L_\Fus} \sum_{i = 1}^{n_a^\A} \alpha_a^i a_i \;\middle|\; a \in L_\Fus, \ 1 \leq i \leq n_a^\A, \ \alpha_a^i \in \mathbb{C}, \ a_i b_j = \sum_{c_k\in L_\A} f_{a_i b_j}^{c_k} c_k \right\}.
\]
Each bimodule \(M\) over \(\A\) has a representation space \(V_M\), whose basis elements are labeled by \(x_i\), where \(x \in L_\Fus\) and \(i = 1, 2, \cdots, n_x^M\), with multiplicities \(n_x^M\). The action of \(\A\) on \(V_M\) is described by a function \(\rho_M: L_\A^2 \times L_M \times L_\Fus \times L_M \to \CC\), representing two algebra elements \((a, b) \in \A^2\) as a rank-3 tensor \([\rho_M]^{ab}_{x_iyz_j}\) on \(V_M\), where set \(L_M\) includes all basis elements \(x_i\). This tensor indicates that \(a\) and \(b\) act sequentially on \(x \in V_M\), transforming it to \(z\in V_M\) with coefficient \(\sum_y [\rho_M]^{ab}_{xyz}\). The intermediate object \(y\) varies within \(L_\Fus\) as per the fusion rule \(\delta_{axy} = \delta_{byz} = 1\).

The duality \(\D\) maps the basic dofs \(M \in L_{\Bimod_\Fus(\A)}\)---simple objects in \(L_{\Bimod_\Fus(\A)}\), which are simple (irreducible) bimodules that cannot be decomposed into direct sums of other bimodules---on each edge and tail of the dual model, to a superposition of the basic dofs on the same edge and tail of the original model:
\eqn{
\D \ \ \Edge{M}\quad :=\quad \frac{1}{d_\A^2}\sum_{a,b \in L_\A} \ \sum_{y\in L_\Fus}\ \sum_{x_i, z_j \in L_{M}} [\rho_M]^{ab}_{x_iyz_j} \ \Charge\ ,\qquad d_\A = \sum_{a\in L_\Fus}n_a^\A d_a,}
where the red lines are auxiliary tails that are to be annihilated by topological moves (See Appendix \ref{sec:pachner}). This linear transformation \(\D\) between the two string-net models is unitary. The multiplicity indices of \(x, z\in L_M\) require additional processing\cite{zhao2024}, but this is not directly relevant to our current discussion and will be addressed in Appendix \ref{sec:generalenlarge}.

\subsection{Mapping Fluxon Condensation to Dyon Condensation}\label{sec:dyoncond}

Now we proceed to construct dyon condensation. To realize a dyon condensation in a given string-net model with input UFC \(\Fus\), we first consider a fluxon condensation in its dual model with input UFC \(\Bimod_\Fus(\A)\) for a certain Frobenius algebra $\A\in\Fus$, given by the projector
\eqn[eq:FluxCondensationTwo]{
P^{\Sub'|\Bimod_\Fus(\A)}_E = \sum_{\text{Dual Condensed Anyons } J} \pi_{J;\Bimod_\Fus(\A)}^{M_0M_0} W_{E;\Bimod_\Fus(\A)}^{J;M_0M_0}\ ,
}
which projects the dofs from varying in \(L_{\Bimod_\Fus(\A)}\) to varying in its subset \(L_{\Sub'} \subset L_{\Bimod_\Fus(\A)}\), where \(M_0\) is the trivial object in \(\Bimod_\Fus(\A)\). The duality map \(\D\) transforms each creation operator \(W_{E;\Bimod_\Fus(\A)}^{J;MN}\) in the dual model into a superposition of creation operators in the original model:
\eqn[eq:DualCreationOperator]{
\D W_{E;\Bimod_\Fus(\A)}^{J;MN}\D^\dagger = \sum_{pq \in L_\Fus} \Delta_J^{MN; pq} W_{E;\Fus}^{J; pq},
}
where \(M_1, M_2\in L_{\Bimod_\Fus(\A)}, \Delta_J^{MN;pq} \in \mathbb{C}\). Since these two equivalent models describe the same output topological order, they share the same set of anyon species\cite{zhao2024}: 
\eq{L_{\Cent(\Fus)} = L_{\Cent(\Bimod_\Fus(\A))}.}
Duality maps transform the basic dofs of models (charges of anyons) but do not alter the topological properties of the order, such as the anyon species \(J\) and their fusion and braiding properties\footnote{The only subtlety is that, since centers are defined based on input UFCs, the same anyon species \(J\) have different names in different models. In certain cases, the dual input UFC is isomorphic to the original UFC by functor \(\mathscr{F}: \Fus\cong\Bimod_\Fus(\A)\). Composing \(\D\) with \(\F\) results in a symmetry transformation within the same original string-net model, which may permute the anyon species. Nevertheless, as we demonstrated in Ref. \cite{zhao2024}, such a symmetry transformation that alters anyon species represents the isomorphism functor \(\mathscr{F}\) rather than the duality map \(\D\), which, in contrast, preserves the anyon species.
}.

Consequently, \(\D\) maps the fluxon condensation projector \(P_E^{\Sub'|\Bimod_\Fus(\A)}\) in the dual model to another projector in the original model:
\eqn[eq:DualProjector]{
P_E^{\Child|\Fus} = \D P_E^{\Sub'|\Bimod_\Fus(\A)} \D^\dagger = \sum_{\text{Dual Condensed Anyons } J} \pi_{J;\Bimod_\Fus(\A)}^{M_0M_0} \Big[\D W_{E;\Bimod_\Fus(\A)}^{J;M_0M_0} \D^\dagger\Big].
}
Knowing the fluxon condensation coefficients \(\pi_{J;\Bimod_\Fus(\A)}^{M_0M_0}\) and the duality transformation coefficients \(\Delta_{JK}^{MN;pq}\), we can directly read out the dyon condensation coefficients \(\pi_J^{pq}\) defined in Eq. \eqref{eq:CondenseHamil} for \(P_E^{\Child|\Fus}\). Projector \(P_E^{\Child|\Fus}\) maps the original model to a child model with Hilbert space and Hamiltonian being
\eq{
\Hil_{\Child} = \Bigg[\prod_{\text{Edges }E}P_E^{\Child|\Fus}\Bigg]\Hil_\Fus,\qquad H_{\Child} = \Bigg[\prod_{\text{Edges }E}P_E^{\Child|\Fus}\Bigg] H_\Fus\Bigg[\prod_{\text{Edges }E}P_E^{\Child|\Fus}\Bigg].
}
The input child UFC \(\Child\) of this child model is isomorphic to \(\Sub' \subset \Bimod_\Fus(\A)\), but now it is a subcategory of the original UFC \(\Fus\). The basic dofs of the child model taking values in \(L_\Child\) are formally labeled by \(\D M\), where \(M \in L_{\Sub'}\). These child basic dofs are expressed as superpositions of the original dofs varying in \(L_\Fus\):
\eqn[eq:GaugeDef]{
\Edge{\D M}\quad :=\quad \frac{1}{d_\A^2}\sum_{a, b \in L_\A} \ \sum_{y\in L_\Fus}\ \sum_{x_i, z_j \in L_{M}} [\rho_M]^{ab}_{x_iyz_j} \quad \Charge\ .
}
The child Hilbert space \(\Hil_{\Child|\Fus}\) is expanded by all possible configurations of the above superposition dofs on each edge and tail, with auxiliary red tails annihilated.

We can extract a categorical description \eqref{eq:pesudo} of our approach, which offers a novel mathematical framework for anyon condensation finer than the traditional UMTC description, which focuses solely on the transformations of topological properties but overlooks the microscopic details during anyon condensation. A fluxon condensation in the equivalent model is mathematically a functor selecting in $\Fus'$ a full subcategory $\Child'$, while a dyon condensation in the original model is a functor selecting in $\Fus$ a subcategory \(\Child\). The duality operator $\D$ that sends the fluxon condensation in the equivalent model to the non-fluxon condensation in the original model is mathematically a \textit{pseudonatural transformation}, which maps parent input $\Fus'$ to $\Fus$, child input UFC $\Child'$ to $\Child$, while the functor selecting $\Child'\subset \Fus'$ to that selecting $\Child\subset \Fus$:
\eqn[eq:pesudo]{\Pseudo\ .}
While this new categorical framework is abstract, our recipe is computable in the string-net model's Hilbert space: Anyons are excited states in the Hilbert space, and duality operators transforming the states are concrete matrices in terms of the model's basic dofs. Therefore, our recipe mathematically represents this finer categorical framework. In particular, a duality operator $\D$ represents a pseudonatural transformation.

The RHS of \eqref{eq:GaugeDef} seems different in lattice structure compared with the original lattice, despite the red lines being auxiliary and to be removed. But the transformed and original lattice structures are the same even before annihilating the red lines for both mathematical and physical reasons. Mathematically, the simple objects in \(L_\Fus\) are simple bimodules over the trivial Frobenius algebra \(\A_0 = \C[1]\); hence 
\eq{
\Fus = \Bimod_{\Fus}(\A_0).
}
Consequently, an edge/tail labeled by simple object \(a\in L_\Fus\) in the original model must also be anchored with two red lines:
\eq{\Edge{a}\qquad \equiv\qquad \TrivialCharge\ .}
Nonetheless, these two red lines are labeled by the trivial object \(1\) and are thus conventionally omitted. Physically, the simple objects of the input UFC of a string-net model are the pure charges, defined concerning the trivial flux characterized by Frobenius algebras. For input UFC \(\Fus\), the trivial flux is characterized by \(\A_0=\C[1]\). For input UFC \(\Bimod_\Fus(\A)\), however, the trivial flux is characterized by \(\A\). The RHS of Eq. \eqref{eq:GaugeDef}, where red lines carry elements of \(\A\), defines precisely how pure charges \(M\in \Bimod_\Fus(\A)\) pertaining to trivial flux \(\A\) appear in the original model with input UFC \(\Fus\).

Let's again consider the Fibonacci string-net model. Our previous work\cite{zhao2024} revealed a Frobenius algebra \(\A\) within \(\Fibo\), where the set of simple objects is \(L_\A = \{1, \tau\}\):
\eq{
\A := \left\{\alpha 1 + \beta\tau \ \middle|\ \alpha, \beta \in \mathbb{C},\ \tau^2 = 1 + \phi^{-\frac{3}{4}} \tau \right\}.
}
Frobenius algebra \(\A\) has two simple bimodules \(M_1\) and \(M_\tau\). The bimodule category \(\Bimod_\Fibo(\A)\) happens to be isomorphic to \(\Fibo\):
\eq{
\Fibo \cong \Bimod_\Fibo(\A),\quad 1 \mapsto M_1,\quad \tau \mapsto M_\tau.
}
As a result, the dual model with input UFC \(\Bimod_\Fibo(\A)\) also has a unique nontrivial fluxon condensation, described by the fluxon condensation term:
\eq{
P_E^{\Sub_0'|\Bimod_\Fibo(\A)} = \frac{\idm + \phi^2 W_E^{M_\tau\bar M_\tau; M_1M_1}}{\phi^2 + 1}.
}
We won't detail the explicit transformation here but directly state the action of the duality map \(\D\) from the dual model to the original model:
\eq{
\D W_{E;\Bimod_\Fibo(\A)}^{M_\tau\bar M_\tau, M_1M_1}\D^\dagger = \frac{1}{\phi^4} W_{E;\Fibo}^{\tau\bar\tau;11} + \frac{\sqrt[4]{5}}{\phi^4} W_{E;\Fibo}^{\tau\bar\tau;\tau 1} + \frac{\sqrt[4]{5}}{\phi^4} W_{E;\Fibo}^{\tau\bar\tau;1\tau} + \frac{\sqrt{5}}{\phi^4} W_{E;\Fibo}^{\tau\bar\tau;\tau\tau}.
}
Therefore, the dyon condensation of the Fibonacci string-net model is defined by the projector
\eqn[eq:FiboDyonProjector]{
P_E^{\Child|\Fibo} = \sum_{J,pq}\frac{\pi_J^{pq}}{d_pd_q}W_E^{J;pq} = \frac{\phi^2\idm + W_{E;\Fibo}^{\tau\bar\tau;11} + \sqrt[4]{5} W_{E;\Fibo}^{\tau\bar\tau;\tau 1} + \sqrt[4]{5} W_{E;\Fibo}^{\tau\bar\tau;1\tau} + \sqrt{5} W_{E;\Fibo}^{\tau\bar\tau;\tau\tau}}{\phi^2(\phi^2 + 1)},
}
where
\eq{\pi_{1\bar 1}^{11} = \frac{1}{\phi^2 + 1},\qquad\ \pi_{\tau\bar\tau}^{11} = \frac{1}{\phi^2(\phi^2 + 1)},\qquad\ \pi_{\tau\bar\tau}^{1\tau} = \pi_{\tau\bar\tau}^{\tau 1} = \frac{\sqrt[4]{5}}{\phi(\phi^2 + 1)},\qquad\ \pi_{\tau\bar\tau}^{\tau\tau} = \frac{1}{\phi}\ .}
One can verify that \(P_E^{\Child|\Fibo}\) is indeed a projector. The child model is also a trivial string-net model with a trivial input UFC, describing a trivial topological order. Each edge/tail possesses a unique trivial dof labeled by \(M_1\) that is more complicated than in the case of fluxon condensations:
\eq{
\Edge{\D M_1} := \sum_{a, b, x, y, z \in \{1, \tau\}} f_{axy}f_{byz} \quad \Charge\ ,
}
where the coefficients \(f_{ijk}\) are given by:
\eq{
f_{111} = f_{1\tau\tau} = f_{\tau 1\tau} = f_{\tau\tau 1} = 1, \qquad f_{\tau\tau\tau} = \frac{1}{\phi^\frac{3}{4}},\qquad f_{11\tau} = f_{\tau 1\tau} = f_{\tau 11} = 0.}
The superposition state on an edge is invariant under the projector \(P_E^{\Child|\Fibo}\) \eqref{eq:FiboDyonProjector}.

\begin{figure}
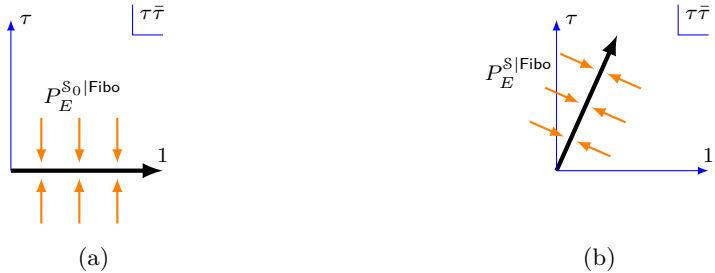
\centering
\subfloat[]{\FibonacciFlux\label{fig:FibonacciA}}\hspace{100pt}
\subfloat[]{\FibonacciDyon\label{fig:FibonacciB}}
\caption{(a) Projection of anyon \(\tau\bar\tau\)'s internal space during fluxon condensation in the Fibonacci string-net model: Projector \(P_E^{\Sub_0|\Fibo}\) projects out dyon \((\tau\bar\tau, \tau)\) but retains fluxon \(\tau\bar\tau, 1\). (b) Projection of the internal space of anyon \(\tau\bar\tau\) during dyon condensation \(P_E^{\Sub|\Fibo}\): A superposition of dyons \((\tau\bar\tau, 1)\) and \((\tau\bar\tau, \tau)\) is condensed.}
\label{fig:Fibonacci}
\end{figure}

After the dyon condensation described by \(P_E^{\Child|\Fibo}\), the new ground state is a coherent state with arbitrarily many \(\tau\bar\tau\) anyons in each plaquette. As we said in Section \ref{sec:condreview}, anyon \(\tau\bar\tau\) splits into two sectors, only one of which condenses. In this coherent state, the condensed sector of anyon \(\tau\bar\tau\) appears as a superposition of dyons \((\tau\bar\tau, 1)\) and \((\tau\bar\tau, \tau)\):
\eqn[eq:ChildCharge]{
\text{Condensed }\tau\bar\tau\text{ Sector} = \frac{1}{\phi^2}(\tau\bar\tau, 1) + \frac{\sqrt[4]{5}}{\phi^2}(\tau\bar\tau, \tau).
}
See Fig. \ref{fig:FibonacciB}. That is, in the internal space of anyon \(\tau\bar\tau\) in the parent Fibonacci string-net model, only the one-dimensional subspace parallel to superposition \eqref{eq:ChildCharge} condenses and becomes the vacuum of the child model. Only one sector of \(\tau\bar\tau\) anyon condenses although both \(W_E^{\tau\bar\tau, 11}\) and \(W_E^{\tau\bar\tau;\tau\tau}\) appears in the condensation term \(P_E^{\Child|\Fus}\).

\section{From Gapping Fundamental DOFs to Condensed Anyons}\label{sec:category}

We have now explicitly established general anyon condensation within the framework of field theory (lattice model), utilizing the language of basic dofs, Hilbert spaces, and Hamiltonians. By gapping out certain dofs on the edges and tails of the model (although these gapped dofs may not directly correspond to simple objects in \(L_\Fus\) but are defined by duality maps), we obtain a child model where the basic dofs take values as simple objects in a subcategory \(\Child\) of the parent UFC \(\Fus\).

On the other hand, in the traditional UMTC categorical description within topological order theory, the condensable boson species of a bosonic anyon condensation in a parent order are described by a CSFA object \(\name{A}\) in the parent UMTC \(\Cent(\Fus)\). A question lies in connecting our model-based framework with this abstract mathematical structure of anyon condensation. We will answer this question at three levels: (1) At the Hamiltonian level, we'll show that the summands \(W_E^{J;pq}\) for condensed anyons of our projector \(P_E^{\Child|\Fus}\) directly represents \(\name{A}\), thereby providing a straightforward algorithm to calculate \(\name{A}\). (2) At the level of basic dofs and input UFC, we'll prove that CSFA \(\name{A}\) is the \emph{full center} of the input child UFC \(\Sub\) within the parent UFC \(\Fus\). (3) At the spectral level, we obtain the creation operators for anyon species in the intermediate and child orders, labeled by simple objects in \(\rep_{\Cent(\Fus)}(\name{A})\) and \(\rep_{\Cent(\Fus), 0}(\name{A})\), respectively.

\subsection{Level 1: Representing and Calculating CSFAs by Projectors}\label{sec:fullcenter}

Let's deal with the first level of our answer. We show that our projector \(P_E^{\Child|\Fus}\), defined as a sum \eqref{eq:CondenseHamil} of creation operators \(W_E^{J;pq}\), is a physical means for directly representing and computing a CSFA \(\name{A}\).

Firstly, let's determine \(\name{A}\)'s basis elements, labeled by \(J_i\), where \(J\) varies in the set of condensed anyon species, which is a subset of \(L_{\Cent(\Fus)}\), and \(i = 1, 2, \cdots, n^{\name{A}}_J\) is the multiplicity label of \(J\). We now use the coefficients \(\pi_J^{pq}\) in projector \(P_E^{\Sub|\Fus}\) \eqref{eq:CondenseHamil} to define the multiplicity \(n^{\name{A}}_J\) and basis elements \(J_i\): Arrange the coefficients \(\pi_J^{pq}\) of a given condensed anyon species \(J\) as a matrix \(\Pi_J\), which is indexed by anyon \(J\)'s charge types \(p_1\) through $p_n$:
\eq{
\Pi_J\quad =\quad \begin{pNiceMatrix}[first-row, first-col]
& p_1 & p_2 & \cdots & p_n \\
p_1 & \pi_J^{p_1p_1} & \pi_J^{p_1p_2} & \cdots & \pi_J^{p_1p_n} \\
p_2 & \pi_J^{p_2p_1} & \pi_J^{p_2p_2} & \cdots & \pi_J^{p_2p_n} \\
\vdots{\ } & \vdots & \vdots & \ddots & \vdots \\
p_n & \pi_J^{p_np_1} & \pi_J^{p_np_2} & \cdots & \pi_J^{p_np_n}
\end{pNiceMatrix}\ .
}
Matrix \(\Pi_J\) is generally not full-rank. We have
\eqn{
n_J^{\name{A}} = {\sf rank}(\Pi_J).
}

We can diagonalize each such matrix \(\Pi_J\) by a similarity transformation \(U^J\), such that diagonalized matrix is indexed by \(\name{A}\)'s basis elements \(J_i\):
\eq{
\Pi_J\quad \to\quad \tilde{\Pi}_J\quad :=\quad U^{J\dagger} \Pi_J U^J\quad =\quad \begin{pNiceArray}[first-row, first-col]{cccc|ccc}
& J_1 & J_2 & \cdots & J_{n_J^\name{A}} & & \cdots & \\
J_1 & \tilde\pi_J^1 & 0 & \cdots & 0 & 0 & \cdots & 0 \\
J_2 & 0 & \tilde\pi_J^2 & \cdots & 0 & 0 & \cdots & 0 \\
\vdots{\ } & \vdots & \vdots & \ddots & \vdots & \vdots & \ddots & \vdots \\
J_{n_J^{\name{A}}} & 0 & 0 & \cdots & \tilde\pi_J^{n_J^{\name{A}}} & 0 & \cdots & 0 \\ \hline
& 0 & 0 & \cdots & 0 & 0 & \cdots & 0 \\
\vdots{\ } & \vdots & \vdots & \ddots & \vdots & \cdots & \ddots & \vdots \\
& 0 & 0 & \cdots & 0 & 0 & \cdots & 0
\end{pNiceArray}.}
Define the corresponding normalized creation operators
\eqn[eq:CondenseCreation]{
\tilde W_E^{J_i} = \frac{d_{\name{A}}\tilde \pi_{J_i}}{d_J}\sum_{J\text{'Charges } p,q} \frac{1}{d_p d_q} (U_J)_{J_i p}^\dagger W_E^{J; pq} (U_J)_{q J_i},
}
which creates two anyons \(J\) with respectively certain superpositions of charge types:
\eqn[eq:conddyon]{
J_i = \sum_{J\text{'s Charges }p}\frac{(U_J)_{J_i p}}{d_p} (J, p).
}
Here,
\eq{d_{\name{A}} = \sum_{\text{Condensed Anyon Species }J}n_J^{\name{A}}d_J}
is the quantum dimension of CSFA \(\name{A}\), where \(d_J\) is the quantum dimension of anyon \(J\) defined in Appendix \ref{sec:spec}.

The condensation projector \(P_E^{\Child|\Fus}\) \eqref{eq:CondenseHamil} is now written as a sum of these normalized creation operators \eqref{eq:CondenseCreation}:
\eq{P_E^{\Child|\Fus} = \frac{1}{d_{\name{A}}}\sum_{J_i\in L_{\name{A}}}d_J\tilde W_E^{J_i},}
which absorbs the creation operators \eqref{eq:CondenseCreation}:
\eqn{P_E^{\Child|\Fus}\tilde W_E^{J_i} = \tilde W_E^{J_i}P_E^{\Child|\Fus} = P_E^{\Child|\Fus},}
where \(L_{\name{A}}\) is the set of all basis elements of \(\name{A}\). Therefore, in the string-net model, we explicitly represent the condensed sectors of the condensed anyon \(J\), which, as mentioned in Section \ref{sec:condreview}, is not captured by the UMTC categorical description. The \(i\)-th condensed sector of \(J\) is \(J_i\) \eqref{eq:conddyon}.

Secondly, we represent the algebra multiplication \(f: L_{\name{A}}^3 \to \mathbb{C}\) of CSFA \(\name{A}\) via operator product expansion of the summands \(\tilde{W}_E^{J_i}\) in projector \(P_E^{\Child|\Fus}\), normalized by quantum-dimension factors:
\eqn[eq:OPE]{
\tilde{W}_E^{J_i} \tilde{W}_E^{K_j} = \sum_{I_k \in L_{\name{A}}}\frac{d_I}{d_Jd_K}f_{J_i K_j}^{I_k}\tilde{W}_E^{I_k}\qquad \forall J_i, K_j\in L_{\name{A}}.
}
The algebra \(\name{A}\) defined as
\eqn[eq:CSFA]{\name{A} = \Bigg\{\sum_{J_i\in L_{\name{A}}}\alpha_{J_i}J_i\Bigg|\alpha_{J_i}\in\CC\Bigg\},\qquad J_iK_j = \sum_{I_k\in L_{\name{A}}}f_{J_iK_j}^{I_k}I_k,\qquad \forall J_i, K_j\in L_{\name{A}}}
is a Frobenius algebra. The brief proof is as follows: In a fusion category \(\Cent(\Fus)\), the Frobenius condition reduces to associativity. This is evident for Eq. \eqref{eq:CSFA}, as the algebra 
\(\name{A}\) is represented as a matrix algebra in Eq. \eqref{eq:OPE}, which is inherently associative. Since duality maps only transform the input UFCs of different models but do not alter the output UMTC that encapsulates the topological properties of the order, any two types of anyon condensation related by a duality map share the same CSFA \(\name{A}\) in the output UMTC. A duality map only transforms the representation of \(\name{A}\) \eqref{eq:OPE}.

Sticking to the example of the Fibonacci string-net model, we present the CSFA of the fluxon condensation and that of the dyon condensation in the model and show that these two CSFAs are identical, agreeing with the known fact in the UMTC description of anyon condensation. 

\begin{enumerate}
\item Fluxon Condensation: Two condensed anyons \(1\bar 1\) and \(\tau\bar\tau\), with \(n_{1\bar 1}^{\name{A}} = n_{\tau\bar\tau}^{\name{A}} = 1\).
\begin{enumerate}
\item Condensed Sector: \(\tau\bar\tau_1 = (\tau\bar\tau, 1)\), a fluxon.
\item Normalized Creation Operators:
\eq{\tilde W_E^{1\bar 1_1} = W_E^{1\bar 1} = \idm,\qquad\qquad \tilde W_E^{\tau\bar\tau_1} = W_E^{\tau\bar\tau;11}.}
\item Algebra multiplication:
\eqn[eq:fiboCSFA]{
f_{1\bar 1_1, 1\bar 1_1}^{1\bar 1_1} = f_{1\bar 1_1, \tau\bar\tau_1}^{\tau\bar\tau_1} = f_{\tau\bar\tau_1, 1\bar 1_1}^{\tau\bar\tau_1} = f_{\tau\bar\tau_1, \tau\bar\tau_1}^{1\bar 1_1} = f_{\tau\bar\tau_1, \tau\bar\tau_1}^{\tau\bar\tau_1} &= 1.
}
\end{enumerate}

\item Dyon condensation: Two condensed anyon species \(1\bar 1\) and \(\tau\bar\tau\), with \(n_{1\bar 1}^{\name{A}} = n_{\tau\bar\tau}^{\name{A}} = 1\).
\begin{enumerate}
\item Condensed Sector: \(\tau\bar{\tau}_1\), a superposition of dyons defined in Eq. \eqref{eq:ChildCharge}:
\eq{
\tau\bar\tau_1 := \frac{1}{\phi^2}(\tau\bar\tau, 1) + \frac{\sqrt[4]{5}}{\phi^2}(\tau\bar\tau, \tau).
}
\item Normalized Creation Operators:
\eq{\tilde W_E^{1\bar 1_1} = \idm,\qquad \tilde W_E^{\tau\bar\tau_1} = \frac{1}{\phi^2}\Bigg(W_E^{\tau\bar\tau;11} + \sqrt[4]{5} W_E^{\tau\bar\tau;\tau 1} + \sqrt[4]{5} W_E^{\tau\bar\tau;1\tau} + \sqrt{5} W_E^{\tau\bar\tau;\tau\tau}\Bigg).
}
\item Algebra multiplication:
\eqn[eq:fiboCSFATwo]{
f_{1\bar 1_1, 1\bar 1_1}^{1\bar 1_1} = f_{1\bar 1_1, \tau\bar\tau_1}^{\tau\bar\tau_1} = f_{\tau\bar\tau_1, 1\bar 1_1}^{\tau\bar\tau_1} = f_{\tau\bar\tau_1, \tau\bar\tau_1}^{1\bar 1_1} = f_{\tau\bar\tau_1, \tau\bar\tau_1}^{\tau\bar\tau_1} &= 1.
}
\end{enumerate}
\end{enumerate}

One can verify that Eqs. \eqref{eq:fiboCSFA} and \eqref{eq:fiboCSFATwo} define the same Frobenius algebra using the definition provided in Appendix \ref{sec:frob} and the categorical data of the doubled Fibonacci topological order listed in Appendix \ref{sec:fibodata}. These two types of anyon condensation are completely indistinguishable within the UMTC description of the doubled Fibonacci topological order.

Thirdly, we prove that the Frobenius algebra \(\name{A}\) \eqref{eq:CSFA} is commutative, meaning that the algebra multiplication \(f_{J_iK_j}^{I_k}\) is commutative: \(f_{J_iK_j}^{I_k} = f_{K_jJ_k}^{I_k}\). Note that commutativity is a property of \(\name{A}\) and does not depend on how the algebra is represented in different models, we only need to show that any such \(\name{A}\) characterizing a fluxon condensation is commutative. The proof is straightforward because fluxon creation operators \(W_E^{J;11}\) are diagonal matrices, rendering the product in Eq. \eqref{eq:OPE} commutative.

Finally, we prove that our constructed commutative Frobenius algebra \(\name{A}\) is separable. The key is that \(\name{A}\) can be faithfully represented by the algebra generated by the creation operators in Eq. \eqref{eq:CSFA}---a subalgebra of the matrix algebra defined over the Hilbert space. Hence, \(\name{A}\) must be semisimple, and a semisimple algebra must be separable.

Therefore, Frobenius algebra \(\name{A}\) \eqref{eq:CSFA} constructed from projector \(P_E^{\Child|\Fus}\) is a CSFA, consistent with the UMTC description of anyon condensation. 

\subsection{Level 2: Full Center}

To understand the categorical relationship between CSFA \(\name{A}\) in \(\Cent(\Fus)\) and the child input UFC \(\Child\), observe that the condensation term \(P_E^{\Child|\Fus}\) is a projector in the parent string-net model. The child dofs \(\D M \in L_{\Child}\) on each edge correspond to those local states on the edge that are invariant under the action of \(P_E^{\Child|\Fus}\):
\eq{P_E^{\Child|\Fus}\quad\Edge{\D M}\quad =\quad \Edge{\D M}\ .}
On the other hand, recall that \(P_E^{\Child|\Fus}\) absorbs condensed sectors' creation operators \(\tilde W_E^{J_i}\) defined in \eqref{eq:CondenseCreation}, the child dofs \(\D M \in L_{\Child}\) on each edge correspond to those local states on the edge that are invariant under the action of \(\tilde W_E^{J_i}\):
\eq{\tilde W_E^{J_i}\quad\Edge{\D M}\quad = \quad\tilde W_E^{J_i}P_E^{\Child|\Fus}\quad\Edge{\D M}\quad =\quad \Edge{\D M}\ .}

The parent output UMTC \(\Cent(\Fus)\) is the center of the parent input UFC \(\Fus\), and the string-net model is a physical representation of this center relationship. The action of creation operator \(\tilde W_E^{J_i}\) embodies the half-braiding of condensed anyon \(J\)'s charges\footnote{The charges appearing in the RHS of Eq. \eqref{eq:conddyon}.} with \(j_E \in \Fus\) on edge \(E\) (see Appendix \ref{sec:halfbraid}). Therefore, the invariance of child basic dofs \(\D M\in L_\Child\) under creation operator \(\tilde W_E^{J_i}\) reflects the categorical fact that \(\name{A}\) \emph{centralizes} the entire subcategory \(\Child\). That is, the half-braiding of any condensed sector \(J_i\) of \(\name{A}\) with any object \(\D M\) in \(\Child\) commutes with all \(\Fus\)'s morphisms involving \(\D M\), so it is trivial. In this context, the CSFA \(\name{A}\) in \(\Cent(\Fus)\) is referred to as the \emph{full center}\cite{fjelstad2008, kong2008, Kong2013} of the child subcategory \(\Child\) within the parent UFC \(\Fus\).

In special cases where the anyon condensation completely breaks the topological order, \(\Child\) is the trivial subcategory containing only the trivial object---the trivial bimodule \(\A\) of the dual input UFC \(\Bimod_\Fus(\A)\). In these instances, the full center \(\name{A}\) of the trivial UFC \(\Sub\) is simply the \emph{center} of the input Frobenius algebra \(\A\)\cite{Hu2018}. Nevertheless, explicitly calculating the center \(\name{A} \in \Cent(\Fus)\) for a given input Frobenius algebra \(\A \in \Fus\) is in general challenging mathematically, not to mention determining the full center of a general input subcategory \(\Sub\) that comprises multiple simple objects that are simple bimodules over \(\A\). Our recipe, in contrast, tackles this challenge from a physical perspective.

\subsection{Level 3: Intermediate Order and child order}\label{sec:rep}

In traditional UMTC categorical description, anyon condensation in a parent order first leads to an auxiliary intermediate order where splitting and identification have occurred. This intermediate order is encapsulated by the representation category \(\rep_{\Cent(\Fus)}(\name{A})\) over the full center \(\name{A}\) in \(\Cent(\Fus)\). Each quasiparticle species \(J_{\rm Int}\) in the intermediate order is labeled by a simple object in \(\rep_{\Cent(\Fus)}(\name{A})\).

In this work, we have not truly let the parameter $\Lambda$ in Hamiltonian \eqref{eq:EffHamil} run to study anyon condensation dynamically. So, we are not sure yet whether such an intermediate order would exist in the phase diagram of our model. Nevertheless, the parent Hilbert space of our model encompasses the possible excited states, i.e., quasiparticles, in the intermediate order. The creation operators of the quasiparticle excitations in the intermediate order can be represented in the parent Hilbert space \(\Hil_\Fus\). Since the intermediate order is simply the child order with the addition of condensed anyons, each creation operator \(W_{E; \rm Int}^{J_{\rm Int}; M_{\rm Int}N_{\rm Int}}\), which creates a pair of \(J_{\rm Int}\) quasiparticles of the intermediate order, can be projected from a corresponding parent creation operator:
\eq{
W_{E; \rm Int}^{J_{\rm Int}; M_{\rm Int}N_{\rm Int}} = W_{E;\Fus}^{J; pq}P_E^{\Child|\Fus},
}
where \(J_{\rm Int}\in L_{\rep_{\Cent(\Fus)}(\name{A})}\), \(P_E^{\Child|\Fus}\) is the anyon condensation projector, and \(W_{E;\Fus}^{J; pq}\) is the creation operator of the parent order, creating a pair of parent anyons \(J\) that become quasiparticle \(J_{\rm Int}\) in the intermediate order. 

To fill the last piece in our puzzle, we should address the explicit relation between an intermediate order and the corresponding child order. This is done by casting the three phenomena caused by anyon condensation listed in Section \ref{sec:condreview}, which are otherwise abstract in the UMTC description, in terms of intermediate creation operators:
\begin{enumerate}
\item An anyon \(J\) splits if its two dyons \((J, p)\) and \((J, q)\) are projected to be distinct intermediate quasiparticle species \(J_{\rm Int}\) and \(J_{\rm Int}'\):
\eq{
W_{E; \rm Int}^{J_{\rm Int}; M_{\rm Int}N_{\rm Int}} =& W_{E;\Fus}^{J; pp}P_E^{\Child|\Fus},\qquad W_{E; \rm Int}^{J'_{\rm Int}; M'_{\rm Int}N'_{\rm Int}} = W_{E;\Fus}^{J; qq}P_E^{\Child|\Fus},\\ 
&W_{E;\Fus}^{J; pq}P_E^{\Child|\Fus} = W_{E;\Fus}^{J; qp}P_E^{\Child|\Fus} = 0.
}
\item Two parent anyons \(J\) and \(J'\) are identified in the intermediate order if their creation operators are projected to be the same:
\eq{W_{E; \rm Int}^{J_{\rm Int}; M_{\rm Int}N_{\rm Int}} = W_{E;\Fus}^{J; pq}P_E^{\Child|\Fus} = W_{E;\Fus}^{J'; p'q'}P_E^{\Child|\Fus}.}
\item An intermediate quasiparticle species \(J_{\rm Int}\) is confined in the child order if its creation operator \(W_E^{J_{\rm Int};MN}\) transforms a child state out of the child Hilbert space \(\Hil_\Child = P_E^{\Child|\Fus}\Hil_\Fus\).
\end{enumerate}
Such confined creation operators are not child creation operators. The true child creation operators, which create a pair of child anyons \(J_\Child \in \rep_{\Cent(\Fus), 0}(\name{A})\), should be:
\eq{
W_E^{J_\Child; M_{\Child}N_{\Child}} = P_E^{\Child|\Fus} W_E^{J;pq} P_E^{\Child|\Fus},
}
where parent anyon \(J\) becomes child anyon \(J_\Child\) after anyon condensation.

\begin{acknowledgments}

We thank Davide Gaiotto, Ling-Yan Hung, Yuting Hu, and Lukas Mueller for inspiring and helpful discussions. YW is supported by the General Program of Science and Technology of Shanghai No. 21ZR1406700, and the Shanghai Municipal Science and Technology Major Project (Grant No. 2019SHZDZX01). The authors are grateful for the hospitality of the Perimeter Institute during his visit, where the main part of this work is done. This research was supported in part by the Perimeter Institute for Theoretical Physics. Research at Perimeter Institute is supported by the Government of Canada through the Department of Innovation, Science and Economic Development and by the Province of Ontario through the Ministry of Research, Innovation and Science. 

\end{acknowledgments}

\appendix

\section{String-net Model}\label{sec:review}

In this section, we briefly review the string-net model defined in Ref. \cite{Hu2018}, which was adapted from that in \cite{Hu2018}. The string-net model is an exactly solvable model defined on a \(2\)-dimensional lattice. An example lattice is depicted in Fig. \ref{fig:lattice}. All vertices are trivalent. Within each plaquette of the lattice, a tail is attached to an arbitrary edge of the plaquette, pointing inward. We will demonstrate that different choices of the edge to which the tail is attached are equivalent in Appendix \ref{sec:pachner}. Each edge and tail is oriented, but we'll show that different choices of directions are equivalent.


The input data of the string-net model is a unitary fusion category \(\Fus\), described by a finite set \(L_\Fus\), whose elements are called \emph{simple objects}, equipped with three functions \(N: L_\Fus^3 \to \NN\), \(d: L_\Fus \to \RR\), and \(G: L^6_\Fus \to \C\). The function \(N\) sets the \emph{fusion rules} of the simple objects, satisfying
\eq{
\sum_{e \in L_\Fus} N_{ab}^e N_{ec}^d = \sum_{e \in L_\Fus} N_{ae}^d N_{bc}^e, \qquad\qquad N_{ab}^c = N_{c^\ast a}^{b^\ast}.
}
There exists a special simple object \(1 \in L_\Fus\), called the \emph{trivial object}, such that for any \(a, b \in L_\Fus\),
\eq{
N_{1a}^b = N_{1b}^a = \delta_{ab},
}
where \(\delta\) is the Kronecker symbol. For each \(a \in L_\Fus\), there exists a unique simple object \(a^\ast \in L_\Fus\), called the \emph{opposite object} of \(a\), such that
\eq{
N_{ab}^1 = N_{ba}^1 = \delta_{ba^\ast}.
}
We only consider the case where for any \(a, b, c \in L_\Fus\), \(N_{ab}^c = 0\) or \(1\). In this case, we define
\eq{
\delta_{abc} = N_{ab}^{c^\ast} \in \{0, 1\}.
}

The basic configuration of the string-net model is established by labeling each edge and tail with a simple object in \(L_\Fus\), subject to the constraint on all vertices that \(\delta_{ijk} = 1\) for the three incident edges or tails meeting at this vertex, all pointing toward the vertex and respectively counterclockwise labeled by \(i, j, k \in L_\Fus\). We can reverse the direction of any edge or tail and simultaneously conjugate its label as \(j \to j^\ast\), which keeps the configuration invariant. The Hilbert space \(\Hil\) of the model is spanned by all possible configurations of these labels on the edges and tails.

The function \(d\) returns the \emph{quantum dimensions} of the simple objects in \(L_\Fus\). It is the largest eigenvalues of the fusion matrix and forms the \(1\)-dimensional representation of the fusion rule.
\eq{d_ad_b = \sum_{c\in L_\Fus}N_{ab}^cd_c.}
In particular, \(d_1 = 1\), and for any \(a\in L_\Fus, d_a = d_{a^\ast}\ne 0\). 

The function \(G\) defines the \(6j\)-\emph{symbols} of the fusion algebra. It satisfies
\eqn[eq:sixj]{\sum_nd_nG^{pqn}_{v^*u^*a}G^{uvn}_{j^*i^*b}G^{ijn}_{q^*p^*c} = G^{abc}_{i^*pu^*}G^{c^*b^*a^*}_{vq^*j},\qquad\sum_n&d_nG^{ijp}_{kln}G^{j^*i^*q}_{l^*k^*n} = \frac{\delta_{pq^*}}{d_p}\delta_{ijp}\delta_{klq},\\
G^{ijm}_{kln} = G^{klm^*}_{ijn^*} = G^{jim}_{lkn^*}= G^{mij}_{nk^*l^*} =& \alpha_m\alpha_n\overline{G^{j^*i^*m^*}_{l^*k^*n^*}},
}
where \(\alpha_a = {\rm sgn}(a)\).

The Hamiltonian of the string-net model reads
\eqn{H := - \sum_{{\rm Plaquettes}\ P}Q_P,\qquad\qquad Q_P := \frac{1}{D}\sum_{s\in L_\Fus}Q_P^s,\qquad D := \sum_{a\in L_\Fus}d_a^2,}
where operator \(Q_P^s\) acts on edges surrounding plaquette \(P\) and has the following matrix elements on a hexagonal plaquette:
\eq{Q_P^s\ \PlaquetteSrc\
:=\ &\delta_{p,0}\ \sum_{j_k\in L_\Fus}\  \prod_{k = 1}^{6}\ \Bigg(\sqrt{d_{i_k}d_{j_k}}\ G^{e_ki_ki_{k+1}^\ast}_{sj_{k+1}^\ast j_k}\Bigg)\PlaquetteTar\ .}
Here, we only show the actions of the \(Q_P\) operator on a hexagonal plaquette. The matrix elements of \(Q_P\) operators on other types of plaquettes are defined similarly.

It turns out that
\eq{
(Q_P^s)^\dagger = Q_P^{s^\ast},\qquad Q_P^rQ_P^s = \sum_{t\in L_\Fus} N_{rs}^tQ_P^t,\qquad Q_P^2 = Q_P,\qquad Q_{P_1}Q_{P_2} = Q_{P_2}Q_{P_1}.
}
The summands \(Q_P\) in Hamiltonian \(H\) are commuting projectors, so the Hamiltonian is exactly solvable. The ground-state subspace \(\Hil_0\) of the system is the projection
\eqn{
\Hil_0 = \Bigg[\prod_{{\rm Plaquettes\ } P}Q_P\Bigg]\Hil.
}
If the lattice has the sphere topology, the model has a unique ground state \(\ket\Phi\) up to scalar factors.

\subsection{Topological Features}\label{sec:pachner}

We briefly review the topological nature of the ground-state subspace of the string-net model defined in Ref. \cite{Hu2018}. Topologically, any two lattices with the same topology can be transformed into each other by so-called \emph{Pachner moves}. There are unitary linear maps between the Hilbert spaces of two string-net models with the same input fusion category on different lattices associated with these Pachner moves, formally denoted as operators \(\T\). The ground states are invariant under such linear transformations. There are three kinds of elementary Pachner moves, whose corresponding linear transformations are:
\eqn[eq:pachner]{
\T \quad \PachnerOne\ ,\\
\T \quad \PachnerTwo\ ,\\
\T \quad \PachnerThree\ .}
Here we use red ``\({\color{red}\times}\)'' to mark the plaquettes to contract. Any other Pachner moves and their corresponding unitary transformations of Hilbert spaces are compositions of these three elementary moves. Given initial and final lattices, there are multiple ways to compose these elementary Pachner moves, but different ways result in the same transformation matrices on the ground-state Hilbert space.

We have also noted that different selections of the edge to which the tail is attached are equivalent. These variations lead to distinct lattice configurations and, consequently, different Hilbert spaces for the lattice model. The equivalence of states in such Hilbert spaces is established by the following linear transformation \(\T'\):
\eqn{
\T'\quad\PachnerFour\ .
}
The states where tails attach to other edges can be obtained recursively in this manner.

For convenience, in certain cases, we will temporarily incorporate auxiliary states with multiple tails within a single plaquette. These states, despite having multiple tails in one plaquette, are all equivalent to states within the Hilbert space:
\eqn{
\PachnerFive\ .
}

\subsection{Excited States}\label{sec:spec}

An \emph{excited state} \(\ket\varphi\) of the string-net model is an eigenstate such that \(Q_P\ket\varphi = 0\) at some plaquettes \(P\). In such a state, we say there are \emph{anyons} in these plaquettes \(P\). We also refer to the ground states as trivial excited states, in which there are only \emph{trivial anyons} in all plaquettes. We assume the sphere topology, in which the model has a unique ground state; nevertheless, the results in this section apply to other topologies.

We start with the simplest excited states with a pair of anyons in two \emph{adjacent} plaquettes with a common edge \(E\). This state can be generated by ribbon operator \(W_E^{J;pq}\):
\eqn{
W_E^{J; pq} \ExcitedA := \sum_{k \in L_\Fus} \sqrt{\frac{d_k}{d_j}} \ \overline{z_{pqj}^{J;k}} \ \ \ExcitedB \ ,
}
where \(j\) is the label on edge \(E\), and \(\bar{z}\) is the complex conjugate. Here, \(z_{pqj}^{J; k}\) is called the \emph{half-braiding tensor} of anyon species \(J\), defined by the following equation:
\eqn[eq:halfA]{
\frac{\delta_{jt}N_{rs}^t}{d_t} z_{pqt}^{J;w} = \sum_{u,l,v\in L_\Fus} z_{lqr}^{J;v} z_{pls}^{J;u} \cdot d_u d_v G^{r^*s^*t}_{p^*wu^*} G^{srj^*}_{qw^*v} G^{s^*ul^*}_{rv^*w}.
}
We will discuss this equation in Appendix \ref{sec:halfbraid}. The label \(J\), called the \emph{anyon species}, labels different minimal solutions \(z^J\) of Eq. \eqref{eq:halfA} that cannot be the sum of any other nonzero solutions. Categorically, anyon species \(J\) are labeled by simple objects in the \emph{center} of UFC \(\Fus\), a modular tensor category whose categorical data record all topological properties of the topological order that the string-net model describes, denoted as \(\Cent(\Fus)\).

The statistics of anyon \(J\) are recorded by its topological spin
\eq{
\theta_{J} = \frac{1}{d_p}\sum_{p \in L_\Fus} d_p z_{ttt}^{J;p},}
where \(t\) is an arbitrary charge of anyon \(J\). The braiding of two anyons \(J\) and \(K\) is recorded by the modular \(S\) matrix, whose matrix elements are
\eq{S_{JK} = \sum_{p, q, k \in L_\Fus} d_k \bar{z}_{ppq}^{J;k} \bar{z}_{qqp}^{K;k}.
}
An anyon \(J\) has trivial self-statistics if \(\theta_J = 1\); two anyons \(J\) and \(K\) braid trivially if and only if \(S_{JK} = d_J d_K\), where \(d_J\) is the quantum dimension of anyon \(J\), defined as
\eq{
d_J = \sum_{J\text{'Charges } p} d_p.
}

States with two quasiparticles in two non-adjacent plaquettes are generated by ribbon operators along longer paths. These longer ribbon operators result from concatenating shorter ribbon operators. For example, to create two quasiparticles \(J^\ast\) and \(J\) with charges \(p^\ast_0\) and \(p_n\) in two non-adjacent plaquettes \(P_0\) and \(P_n\), we can choose a sequence of plaquettes \((P_0, P_1, \cdots, P_n)\), where \(P_i\) and \(P_{i+1}\) are adjacent plaquettes with their common edge \(E_i\). The ribbon operator \(W_{P_0P_n}^{J; p_0p_n}\) is
\eq{
W_{P_0P_n}^{J; p_0p_n} := \left[\sum_{p_1 p_2 \cdots p_{n-1} \in L_\Fus} \prod_{k = 1}^{n-1} \left(d_{p_k} B_{P_k} W_{E_k}^{J; p_k p_{k+1}}\right)\right] W_{E_0}^{J; p_0 p_1}.
}
Different choices of plaquette paths \((P_0, P_1, \cdots, P_n)\) give the same operator \(W_{P_0P_n}^{J; p_0 p_n}\) if these sequences can deform continuously from one to another. Following the same procedure, we can also define the creation operator of three or more anyons.

At the end of this section, we define the measurement operator \(\Pi_P^J\) measuring whether there is an anyon \(J\) excited in plaquette \(P\):
\eqn{\Pi_P^J\ \PlaquetteSrc\qquad
:=\qquad \sum_{s, t\in L_\Fus}\ \frac{d_sd_t}{d_p}z_{pps}^{J; t}\quad \PlaquetteMsr\quad.}
The set of measurement operators are orthonormal and complete:
\eq{\Pi_P^J\Pi_P^K = \delta_{JK}\Pi_P^J,\qquad \sum_{J\in L_{\mathcal{Z}(\Fus)}}\Pi_P^J = \idm.}

\subsection{The Output UMTC is the Center of the Input UFC}\label{sec:halfbraid}

As mentioned earlier, the string-net model's output UMTC \(\Cent(\Fus)\) is the center of its input UFC \(\Fus\), and it is a physical representation of this center relationship. In this appendix, we explicitly demonstrate how this representation is understood.

Categorically, an object \(J\) in center \(\Cent(\Fus)\) is denoted as a pair \(J = (X_J, c_{X_J, \cdot})\), where \(X_J\) is an object in UFC \(\Fus\), and \(c_{x_J, \cdot}\) is called a \emph{half-braiding}, which is a set of morphisms
\eq{
c_{X_J, y}: X_J\otimes y\to y\otimes X_J.
}
A morphism \(c_{X_J, y}\) braids object \(X_J\) with object \(y\) in \(\Fus\) and can be depicted as
\eq{\HalfBraidingA.}
In fusion category \(\Fus\), all morphisms can be decomposed as direct sums of fusion of simple objects, and so can the half-braiding:
\eqn[eq:halfB]{\HalfBraidingB.}
Here, \(L_J\subseteq L_\Fus\), such that the direct sum of simple objects in \(L_J\) is \(X_J\):
\eq{X_J = \bigoplus_{p \in L_J} p.}
The expansion coefficients \(z_{pqk}^{J;y}\) are known as the \emph{half-braiding tensor} of \(J\). A half-braiding should commute with any fusion in \(\Fus\):
\eqn[eq:halfC]{\HalfBraidingC,\qquad \forall p, q\in L_J,\quad y, u, v\in L_\Fus.}
Expanding Eqs. \eqref{eq:halfC} using Eq. \eqref{eq:halfB} leads to Eq. \eqref{eq:halfA}.

For a string-net model with input UFC $\Fus$, an anyon type \(J\) is a simple object in the output UMTC \(\Cent(\Fus)\), and $J$'s charges take value in \(L_J\). The action of creation operator \(W_E^{J;pq}\) directly represents the half-braiding morphism \(c_{X_J, j_E}\) of object \(X_J\) with \(j_E\in L_\Fus\), the dof on edge \(E\):
\eq{\HalfBraidingD}

\section{Duality Maps of String-Net Model}\label{sec:symmtrans}

In this section, we briefly review the duality maps of the string-net models constructed in Ref. \cite{zhao2024}.

It is a mathematical theorem \cite{etingof2016} that two fusion categories \(\Fus\) and \(\Fus'\) have isomorphic centers if and only if they are \emph{categorically Morita equivalent}. That is, two string-net models with categorically Morita equivalent input fusion categories describe the same topological order. Category theory also tells that if a fusion category \(\Fus'\) is categorically Morita equivalent to \(\Fus\), there must be a \emph{Frobenius algebra} \(\A\) in \(\Fus\), such that \(\Fus'\) is isomorphic to the \emph{bimodule category} over \(\A\) in \(\Fus\):
\eqn{
\Fus' \cong \Bimod_\Fus(\A).
}
Therefore, different string-net models describing the same topological order are classified by all Frobenius algebras \(\A\) in a particular input fusion category \(\Fus\). Such equivalent models have bimodule categories \(\Bimod_\Fus(\A)\) as their input fusion categories. We can establish the duality maps between these equivalent models. In this section, we briefly review the definition of Frobenius algebras in a given fusion category and their bimodules and leave the duality maps for the next section.

\subsection{Frobenius Algebra}\label{sec:frob}

A Frobenius algebra \(\A\) in a fusion category \(\Fus\) is characterized by a pair of functions \((n, f)\).  Function \(n: L_\Fus \to \NN\) returns the \emph{multiplicity} \(n_a\) of \(a\in L_\Fus\) appearing in the Frobenius algebra \(\A\), satisfying \(n_a = n_{a^\ast}\). The basis elements of algebra \(\A\) are labeled by \(a_\alpha\), where \(a \in L_\Fus\) satisfies \(n_a > 0\), and \(\alpha = 1, 2, \ldots, n_a\) is the \emph{multiplicity index}. We denote the set of all basis elements in \(\A\) as \(L_\A\). 

The algebra multiplication of \(\A\) is given by function \(f: L^3_\A \to \C\), satisfying:
\eqn[eq:frob]{\sum_{t_\tau \in L_\A} f_{r_\rho s_\sigma t_\tau} f_{a_\alpha b_\beta t_\tau^\ast} G^{rst}_{abc} \sqrt{d_c d_t} &= \sum_{\gamma = 1}^{n_c} f_{a_\alpha c_\gamma s_\sigma} f_{r_\rho c_\gamma^\ast b_\beta}\ ,\\ \\
\sum_{a_\alpha b_\beta \in L_\A} f_{a_\alpha b_\beta c_\gamma} f_{b_\beta^\ast a_\alpha^\ast c_\gamma^\ast} \sqrt{d_a d_b} = d_\A \sqrt{d_c},\qquad f_{a_\alpha b_\beta c_\gamma} &= f_{b_\beta c_\gamma a_\alpha}, \qquad f_{0 a_\alpha b_\beta} = \delta_{ab^\ast} \delta_{\alpha\beta},}
where
\eqn{
d_\A := \sum_{a \in L_\Fus} n_a d_a
}
is the \emph{quantum dimension} of \(\A\). This definition aligns with the one in the main body, where a Frobenius algebra \(\A\) is expressed as a vector space spanned by basis elements in \(L_\A\), and the algebraic multiplicity rule is given by function \(f\):
\eq{
\A = \C[L_\A],\qquad a_\alpha b_\beta = \sum_{c_\gamma \in L_\A} f_{a_\alpha b_\beta c_\gamma^\ast} \, c_\gamma =  f_{a_\alpha b_\beta}^{c_\gamma} \, c_\gamma \in \C[L_\A].
}

For convenience, in a lattice model, we use red edges or tails to indicate that this edge or tail is labeled by a basis element in Frobenius algebra \(\A\), and a red dot on a vertex to represent a coefficient \(f\) multiplied to this state. Graphically,
\eqn{
\FrobeniusA\ .
}
We also use dashed red edges or tails to represent that we are summing over all states with labels on this edge in \(L_\A\). For example, the definition \eqref{eq:frob} of Frobenius algebra \(\A\) can then be illustrated graphically by the Pachner moves of string-net models:
\eq{\FrobeniusB\ ,}
\eq{\FrobeniusC\ .}

\subsection{Bimodules over a Frobenius Algebra}\label{sec:bimod}

A bimodule \(M\) over a Frobenius algebra \(\A\) in a fusion category \(\Fus\) is characterized by a pair of functions \((n^M, P_M)\). The function \(n^M: L_\Fus \to \NN\) returns the \emph{multiplicity} \(n^M_a\) of \(a \in L_\Fus\) appearing in bimodule \(M\), satisfying \(n^M_a = n^M_{a^\ast}\). The basis elements of \(M\) are labeled by pairs \(a_i\), where \(a \in L_\Fus\) satisfies \(n_a^M > 0\), and \(i = 1, 2, \ldots, n^M_a\) labels the multiplicity index. We denote the set of all basis elements in bimodule \(M\) as \(L_M\). 

The action of Frobenius algebra \(\A\) on bimodule \(M\) is characterized by function \(P_M: L_\A^2 \times L_M \times L_\Fus \times L_M \to \C\), satisfying the following defining equations:
\eqn[eq:bimod]{&\sum_{uv\in L_\Fus}\ \sum_{y_\upsilon\in L_M}\ [P_M]^{a_\alpha r_\rho}_{x_\chi u y_\upsilon}\ [P_M]^{b_\beta s_\sigma}_{y_\upsilon v z_\zeta}\ G^{v^\ast by}_{urw}\ G^{w^\ast bu}_{axc}\ G^{sz^\ast v}_{wrt^\ast}\ \sqrt{d_ud_vd_wd_yd_cd_t}\\ 
=\ &\sum_{\gamma = 1}^{n_c}\ \sum_{\tau = 1}^{n_t}\ P^{c_\gamma t_\tau}_{x_\chi w z_\zeta}f_{a_\alpha c_\gamma^\ast b_\beta}\ f_{r_\rho s_\sigma t_\tau},\\
&\qquad\qquad [P_M]^{00}_{x_\chi y z_\zeta} = \delta_{xy}\delta_{yz}\delta_{\chi\upsilon}\delta_{\upsilon\zeta},\qquad [P_M]^{a_\alpha b_\beta}_{x_\chi y z_\zeta} = [P_M]^{b_\beta a_\alpha}_{z_\zeta^\ast y^\ast x_\chi^\ast}.
}

In the string-net model, we use blue lines to represent that this edge is labeled by a basis element in \(L_M\), and blue dashed line to represent that we are summing over all intermediate labels in \(L_\Fus\) with coefficients \(P_M\):
\eq{\BimoduleA\ .}
This definition \eqref{eq:bimod} of bimodule \(M\) can then be depicted graphically by Pachner moves:
\eq{
\BimoduleB\ .
}
This definition aligns with the one in the main body, where a bimodule \(M\) is expressed as a vector space spanned by basis elements in \(L_M\). A pair of Frobenius algebra elements \((a_\alpha, b_\beta)\in \C[L_\A]^2\) is represented as a three-index tensor \(P_M\) on the bimodule space \(\C[L_M]\).

\subsection{General Constructions of Dualities and Symmetry Transformations in the Extended String-Net Model}\label{sec:trans}

Given a fusion category \(\Fus\) and a Frobenius algebra \(\A \in \Fus\), two string-net models with \(\Fus\) and \(\Bimod_\Fus(\A)\) as the input data describe the same topological order. Categorically, \(\Bimod_\Fus(\A)\) is defined by an injective functor
\eqn{
\D: \Bimod_\Fus(\A) \to \Fus, \qquad M \mapsto \bigoplus_{a \in L_\Fus} n^M_a a,
}
and for any morphisms \(\phi_{M_1 M_2}^{M_3} \in \Bimod_\Fus(\A): M_1 \otimes M_2 \to M_3\) and \(\varphi_{xy}^z \in \Fus: x \otimes y \to z\),
\eq{
\D(\phi_{M_1 M_2}^{M_3}) = \bigoplus_{z_\zeta \in L_{M_3}} \Bigg[ \bigoplus_{x_\chi \in L_{M_1}} \bigoplus_{y_\upsilon \in L_{M_2}} \mathcal{V}_{M_1 M_2 M_3^\ast}^{x_\chi y_\upsilon z_\zeta^\ast} \ \varphi_{x_\chi y_\upsilon}^{z_\zeta} \Bigg],
}
where \(x_\chi\), \(y_\upsilon\), and \(z_\zeta\) are respectively the \(\chi\)-th \(x\) object,  \(\upsilon\)-th \(y\), and \(\zeta\)-th \(z\) in \(\D(M)\).

Such a functor \(\D\) induces a duality map between the Hilbert spaces of two string-net models with \(\F\) and \(\Bimod_\Fus(\A)\) as the input data. On each edge, \(D\) transforms the basic dofs as
\eqn{
\D\qquad\Edge{M} \qquad =\qquad \sum_{a_\alpha, b_\beta\in L_\A}\sum_{x_\chi, z_\zeta\in L_M}\BimoduleG\ .
}
This duality induces a unitary morphism between the Hilbert spaces \(\Hil_{\Bimod_\Fus(\A)}\) and \(\Hil_{\Fus}\) of these two models, which can be understood plaquette by plaquette:
\begin{align}
\BimoduleH\ .\label{eq:pladuality}
\end{align}
Note that the black edges and tails labeled by \(I_i, E_i, M \in L_{\Bimod_\Fus(\A)}\) represent basis states in the dual model, where \(\Bimod_\Fus(\A)\) is the input fusion category and \(I_i, E_i, M\) are simple objects. In contrast, the blue edges and tails labeled by \(I_i, E_i, M \in \Bimod_\Fus(\A)\) represent superposition states in the original model with \(\Fus\) as the input fusion category. \(\D\) is a unitary map up to a global scalar coefficient detailed in Ref. \cite{zhao2024}.

After the topological moves in Eq. \eqref{eq:pladuality}, the dof on any edge will cease to have any multiplicity index of simple objects in bimodules, while that on any tail will still have a multiplicity index: 
\eqn[eq:FiboDualPlaq]{
\FibonacciC,
}
where \(I_i, E_i, M \in L_{\Bimod_\Fus(\A)}\), and ``\(\cdots\)'' omits the expansion coefficients after topological moves.
Therefore, to make sense of this duality and make it unitary, we are urged to enlarge the Hilbert space of the original Fibonacci string-net model on each tail but not on the edges, such that two simple objects \(a_\alpha, a_\beta \in L_M\) with different multiplicity indices \(\alpha\ne\beta\) are distinguishable on tails. This enlargement is physically sound: The tail carries an anyon's internal charge that reflects the action of \(\A\), which can only be told when different occurrences of the same \(a\) in the bimodules of \(\A\) are distinguished by multiplicity indices as \(a_i\). In contrast, the dofs on edges are about ground states because any path along edges has to be a closed loop. At any vertex along such a loop, fusion rules are met; they treat two simple objects \(a_\alpha, a_\beta \in L_M\) with different multiplicity indices \(\alpha\ne\beta\) the same\footnote{As an analogy: It makes no sense to question the electric charge in a closed electric flux loop because the Gauss law (analogous to fusion rules) is met everywhere along the loop. Only when the loop is cut open to be a path, one can ask about the charges at the ends of the path where the Gauss law is broken.}. 

\subsection{Enlarging the Hilbert Space}\label{sec:generalenlarge}

In the enlarged Hilbert space, each tail carries a dof labeled by a pair \(a_\alpha\), where 
\eqn{a\in L_\mathscr{F},\qquad \alpha = 1, 2, \cdots, N^\mathcal{A}_a,\qquad N^\mathcal{A}_a = \max_{M\in L_{{\rm Bimod}_\mathscr{F}}(\mathcal{A})}\{n^M_a\},
}
where \(L_{\Bimod_\F(\A)}\) is the set of all simple bimodules over Frobenius algebra \(\A\). But the basic dofs on edges remain to take value varying the simple objects of the input fusion category \(\Fus\). The Hilbert space on the tail is spanned by all enlarged dofs on tails and original dofs on edges, subject to the fusion rules on all vertices. 

For any bimodule \(M\), its simple object \(x^M_\chi\in L_M\) corresponds to a superposition state \(\ket{x^M_\chi}\) in the local Hilbert space of a tail:
\eqn{
\ket{x^M_\chi} := \sum_{i = 1}^{N^\mathcal{A}_x} A^{x,M}_{\chi,i}\ket{x_i}.
}
All different states should satisfy the orthonormal conditions:
\eqn{\BimoduleF\ .}

\section{Fibonacci Fusion Category and Frobenius Algebra}

In this section, we list the categorical data of the Fibonacci UFC and the doubled Fibonacci UMTC. 

\subsection{Categorical Data of the Fibonacci UFC}
The Fibonacci fusion category \(\Fibo\) has two simple objects, denoted as \(1\) and \(\tau\). The nonzero fusion rules are \eq{\delta_{111} = \delta_{1\tau\tau} = \delta_{\tau\tau\tau} = 1,}
and the quantum dimensions are 
\eq{d_1 = 1,\qquad d_\tau = \phi = \frac{\sqrt{5} + 1}{2}.}
The nonzero independent \(6j\) symbols are
\eq{G^{111}_{111} = 1,\qquad G^{111}_{\tau\tau\tau} = \frac{1}{\sqrt{\phi}},\qquad G^{1\tau\tau}_{1\tau\tau} = G^{1\tau\tau}_{\tau\tau\tau} = \frac{1}{\phi},\qquad G^{\tau\tau\tau}_{\tau\tau\tau} = -\frac{1}{\phi^2}.}

\subsection{Simple Bimodules over the Frobenius Algebra in Fibonacci UFC}\label{sec:fibodata}

Fibonacci fusion category \(\Fibo\) has a nontrivial Frobenius algebra \(\A\), such that
\eq{L_\mathcal{A} = \{1, \tau\},\qquad f_{111} = f_{1\tau\tau} = f_{\tau 1\tau} = f_{\tau\tau 1} = 1,\qquad f_{\tau\tau\tau} = -\frac{1}{\phi^\frac{3}{4}}.}
There are two simple bimodules over \(\A\), denoted as \(M_1\) and \(M_\tau\), such that
\eq{L_{M_1} = \{1, \tau\},\qquad [P_{M_1}]^{ab}_{xyz} = f_{axy}f_{byz},\qquad L_{M_\tau} = \{1, \tau_1, \tau_2\},}
\eq{[P_{M_\tau}]^{11}_{111} = [P_{M_\tau}]^{11}_{\tau_0\tau\tau_0} = [P_{M_\tau}]^{11}_{\tau_1\tau\tau_1} = 1\ ,}
\eq{[P_{M_\tau}]^{1\tau}_{11\tau_0} = [P_{M_\tau}]^{1\tau}_{\tau_1\tau 1} = [P_{M_\tau}]^{\tau 1}_{\tau_011} = [P_{M_\tau}]^{\tau 1}_{1\tau\tau_1} = \frac{1}{2\phi} + \frac{\sqrt{\phi}}{2}i\ ,}
\eq{[P_{M_\tau}]^{1\tau}_{11\tau_1} = [P_{M_\tau}]^{1\tau}_{\tau_0\tau 1} = [P_{M_\tau}]^{\tau 1}_{\tau_111} = [P_{M_\tau}]^{\tau 1}_{1\tau\tau_0} = \frac{1}{2\phi} - \frac{\sqrt{\phi}}{2}i\ ,}
\eq{[P_{M_\tau}]^{1\tau}_{\tau_0\tau\tau_0} = [P_{M_\tau}]^{1\tau}_{\tau_1\tau\tau_1} = [P_{M_\tau}]^{\tau 1}_{\tau_0\tau\tau_0} = [P_{M_\tau}]^{\tau 1}_{\tau_1\tau\tau_1} = -\frac{\sqrt[4]{\phi}}{2\phi^2}\ ,}
\eq{[P_{M_\tau}]^{1\tau}_{\tau_0\tau\tau_1} = [P_{M_\tau}]^{\tau 1}_{\tau_1\tau\tau_0} = -\frac{\sqrt[4]{\phi}}{2} - \frac{\phi^\frac{3}{4}}{2}i\ ,\qquad [P_{M_\tau}]^{1\tau}_{\tau_1\tau\tau_0} = [P_{M_\tau}]^{\tau 1}_{\tau_0\tau\tau_1} = -\frac{\sqrt[4]{\phi}}{2} + \frac{\phi^\frac{3}{4}}{2}i\ ,}
\eq{[P_{M_\tau}]^{\tau\tau}_{1\tau1} = -\frac{1}{\phi}\ ,\qquad [P_{M_\tau}]^{\tau\tau}_{1\tau\tau_0} =[P_{M_\tau}]^{\tau\tau}_{1\tau\tau_1} =[P_{M_\tau}]^{\tau\tau}_{\tau_0\tau 1} =[P_{M_\tau}]^{\tau\tau}_{\tau_1\tau 1} = -\frac{\sqrt[4]\phi}{\phi}\ ,}
\eq{[P_{M_\tau}]^{\tau\tau}_{\tau_01\tau_0} = -\frac{1}{2\phi} + \frac{i}{2\sqrt{\phi}}\ ,\quad\ [P_{M_\tau}]^{\tau\tau}_{\tau_11\tau_1} = -\frac{1}{2\phi} - \frac{i}{2\sqrt{\phi}}\ ,\quad\ [P_{M_\tau}]^{\tau\tau}_{\tau_01\tau_1} = [P_{M_\tau}]^{\tau\tau}_{\tau_11\tau_0} = \frac{1}{2}\ ,}
\eq{[P_{M_\tau}]^{\tau\tau}_{\tau_0\tau\tau_1} = [P_{M_\tau}]^{\tau\tau}_{\tau_1\tau\tau_0} = \frac{\sqrt{\phi}}{2\phi^2}\ ,\quad\ [P_{M_\tau}]^{\tau\tau}_{\tau_0\tau\tau_0} = -\frac{\sqrt{\phi}}{2\phi^3} - \frac{\phi}{2}i\ ,\quad\ [P_{M_\tau}]^{\tau\tau}_{\tau_1\tau\tau_1} = -\frac{\sqrt{\phi}}{2\phi^3} + \frac{\phi}{2}i\ .}

The bimodule category \(\Bimod_\Fibo(\A)\), a UFC with two simple objects  \(M_1, M_\tau\), has the following categorical data:
\eq{d_{M_1} = 1,\qquad d_{M_\tau} = \phi,\qquad \delta_{M_1M_1M_1} = \delta_{M_1M_\tau M_\tau} = \delta_{M_\tau M_\tau M_\tau} = 1,\qquad G^{M_1M_1M_1}_{M_1M_1M_1} = 1,}
\eq{G^{M_1M_1M_1}_{M_\tau M_\tau M_\tau} = \frac{1}{\sqrt{\phi}},\qquad G^{M_1M_\tau M_\tau}_{M_1M_\tau M_\tau} = G^{M_1M_\tau M_\tau}_{M_\tau M_\tau M_\tau} = \frac{1}{\phi},\qquad G^{M_\tau M_\tau M_\tau}_{M_\tau M_\tau M_\tau} = -\frac{1}{\phi^2}.}
Evidently, \(\Bimod_\Fibo(\A)\) is isomorphic to UFC \(\Fibo\) by functor
\eq{\F_\A: \Fus \to \Bimod_\Fus(\A),\qquad 1\mapsto M_1,\qquad \tau\mapsto M_\tau.}

\subsection{Categorical Data of Doubled Fibonacci UMTC}

The doubled Fibonacci UMTC \(\Cent(\Fibo)\) has four simple objects:
\eq{1\bar 1,\qquad 1\bar\tau,\qquad \tau\bar 1,\qquad \tau\bar\tau,}
whose corresponding half-braiding tensors are
\eq{z_{111}^{1\bar 1; 1} = z_{11\tau}^{1\bar 1; \tau} = 1;}
\eq{z_{\tau\tau 1}^{\tau\bar 1; \tau} = 1,\qquad z_{\tau\tau\tau}^{\tau\bar 1; 1} = -\frac{\phi}{2} - \frac{i}{2}\sqrt{\frac{\sqrt{5}}{\phi}},\qquad z_{\tau\tau\tau}^{\tau\bar 1; \tau} = -\frac{1}{2\phi} + \frac{i}{2}\sqrt{\sqrt{5}\phi};}
\eq{z_{\tau\tau 1}^{1\bar\tau; \tau} = 1,\qquad z_{\tau\tau\tau}^{\tau\bar 1; 1} = -\frac{\phi}{2} + \frac{i}{2}\sqrt{\frac{\sqrt{5}}{\phi}},\qquad z_{\tau\tau\tau}^{\tau\bar 1; \tau} = -\frac{1}{2\phi} - \frac{i}{2}\sqrt{\sqrt{5}\phi};}
\eq{z_{111}^{\tau\bar\tau; 1} = 1,\qquad z_{11\tau}^{\tau\bar\tau; \tau} = -\frac{1}{\phi^2},&\qquad z_{\tau\tau 1}^{\tau\bar\tau; \tau} = 1,\qquad z_{\tau\tau\tau}^{\tau\bar\tau; 1} = 1,\qquad z_{\tau\tau\tau}^{\tau\bar\tau; \tau} = \frac{1}{\phi^2},\\
&z_{1\tau\tau}^{\tau\bar\tau; \tau} = z_{\tau 1\tau}^{\tau\bar\tau; \tau} = \pm\frac{\sqrt[4]{5}}{\phi}.}

Note that each simple object in \(\Cent(\Fibo)\) is formally written in the form \(a\bar a'\), where \(a,a'\in\{1, \tau\}\), the fusion categorical property of \(\Cent(\Fibo)\) is
\eq{d_{a\bar a'} = d_ad_{a'},\qquad \delta_{a\bar a', b\bar b', c\bar c'} = \delta_{abc}\delta_{a'b'c'},\qquad G^{a\bar a', b\bar b', m\bar m'}_{c\bar c', d\bar d', n\bar n'} = G^{abm}_{cdn}\bar G^{a'b'm'}_{c'd'n'}.}
The braiding properties of simple objects in UMTC \(\Cent(\Fibo)\) is recorded in the modular \(S\) and \(T\) matrices of \(\Cent(\Fibo)\):
eq{S = ,\qquad T = }
There is a unique nontrivial CSFA \(\name{A}\) in UMTC \(\Cent(\Fibo)\), such that
\eq{L_{\name{A}} = \{1\bar 1, \tau\bar\tau\},\qquad f_{1\bar 1, 1\bar 1, 1\bar 1} = f_{1\bar 1, \tau\bar\tau, \tau\bar\tau} = f_{\tau\bar\tau, \tau\bar\tau, \tau\bar\tau} = 1.}

\bibliographystyle{apsrev4-1}
\bibliography{StringNet}
\end{document}